\documentclass[11 pt]{iopart}
\usepackage{iopams,mathrsfs,graphicx,hyperref,color}
\expandafter\let\csname equation*\endcsname\relax
\expandafter\let\csname endequation*\endcsname\relax
\newcommand{\gae}{\lower 2pt \hbox{$\, \buildrel {\scriptstyle >}\over {\scriptstyle
\sim}\,$}}
\newcommand{\lae}{\lower 2pt \hbox{$\, \buildrel {\scriptstyle <}\over {\scriptstyle
\sim}\,$}}
\usepackage{amsmath,relsize,ulem}

\def\i{\ket{\psi (0)}}

\def\ket#1{\mathinner{|{#1}\rangle}}

\begin{document}
%%%%%%%%%%%%%%%%%%%%%%%%%%%%%%%%%%%%%%%%%%%%%%%%%%%%%%%%%%%%%%%%%%%%%%%%%%%%%%%
\title{Dynamics and steady states of tight-binding chains in presence of isolated defects}
\author{Anish Acharya}
\address{Department of Theoretical Physics, Tata Institute of Fundamental Research, Homi Bhabha Road, Mumbai 400005, India}
\ead{anish.acharya@tifr.res.in}
\author{Luca Giuggioli}
\address{School of Engineering Mathematics and Technology, University of Bristol, Bristol BS8 1TW, United Kingdom}
\ead{luca.giuggioli@bristol.ac.uk}
\author{Shamik Gupta}
\address{Department of Theoretical Physics, Tata Institute of Fundamental Research, Homi Bhabha Road, Mumbai 400005, India}
\ead{shamikg1@gmail.com}

\begin{abstract}
Reduced transport and localization in isolated quantum systems are typically attributed to spatially-extended disorder, but may also emerge from the influence of a few controllable defects. We show here how a single defect profoundly reshapes wave-function spreading on a finite and periodic tight-binding lattice. Adapting the defect technique from classical random-walk studies, we obtain exact time-resolved site-occupation probabilities and several observables of interest. Even a single defect induces remarkable nonlinear effects, including non-monotonic suppression of transport, enhanced localization at distant sites, and strong sensitivity to the initial particle position at long times. These results demonstrate that minimal perturbations can generate nontrivial long-time transport signatures, giving rise to a microscopic defect-driven mechanism of quantum localization. Although the main results presented pertain to a single isolated defect, we show that the developed formalism may naturally extend to multiple as well as to a wider class of defects.
\end{abstract}
\maketitle

%\tableofcontents
%%%%%%%%%%%%%%%%%%%%%%%%%%%%%%%%%%%%%%%%%%%%%%%%%%%%%%%%%%%%%%%%%%%%%%%%%%%%%%%

\section{Introduction}
\label{sec1:intro}
Understanding the dynamics of isolated quantum systems remains a central pursuit in condensed matter and statistical physics~\cite{RevModPhys.83.863,AdvPhysics2016,RevModPhys.91.021001,Mori_2018,nandkishore2015many,essler2016quench}. Disorder can dramatically modify the dynamics, e.g., in lattice systems, uncorrelated site-wise disorder induces destructive interference that inhibits spreading of wave functions, leading to fundamental phenomena such as Anderson localization~\cite{PhysRev.109.1492}. Localization can also be induced dynamically, e.g., through the application of an external periodic field, and can even be prevented through a careful choice of the field amplitude and frequency~\cite{PhysRevB.34.3625,DUNLAP1988438,PhysRevB.37.6622}. Even in the absence of true spatial randomness, quasi-periodic potentials, as in the Aubry–Andr\'{e} model~\cite{Aubry1980}, can result in either an extended or a localized phase, with a phase transition between the two on tuning the amplitude of the potential.

In the literature, the majority of theoretical and experimental work has focused on spatially-extended disorder, wherein randomness permeates the entire system and governs transport and localization properties \cite{annurev@Vojta,nonh_transport@Zhong2025}. By contrast, localized defects, namely, heterogeneities restricted to a few lattice sites, arise naturally in finite engineered platforms such as cold-atom and photonic lattices \cite{DynLocExp2006,Iyer2007ExactDL}. While established frameworks, including scaling theories of localization \cite{Abrahams1979}, flow-equation methods \cite{thomson2020dynamics}, and strong-disorder renormalization-group approaches \cite{Igloi2005}, have elucidated the role of spatially-extended disorder, the dynamical consequences of localized defects in finite isolated systems also play an important role in elucidating the effects of spatial heterogeneities on the transport properties of a quantum system~\cite{igloi2014evolution}. Here, we consider a specific aspect of this broader topic and provide a focused theoretical analysis within this setting~\cite{PhysRevLett.103.080404}. Our approach relies on the Schr\"{o}dinger time evolution of the wave function of the system, rather than the density-matrix master equation~\cite{hakenreineker1972,Reineker1991}, transfer-matrix method~\cite{PhysRev.131.1500}, or variational theory frameworks~\cite{verma1971} adopted in earlier works~\cite{maraddudin1971theory}.

In this work, we extend to the quantum domain the defect technique, developed originally in classical transport theory for spatially-disordered systems~\cite{Montroll1967,Montroll1969,montroll1975scattering,kenkre2021memory}, and examine analytically how isolated heterogeneities influence the dynamics of  quantum systems. By exploiting the power of the defect technique, e.g., as demonstrated recently in the discovery of the so-called disorder mean indifference phenomenon in classical random walk first-passage studies~\cite{PhysRevResearch.5.043281,Barbini_2024,Giuggioli_2024}, we develop an analytically-exact formalism to unveil the spatio-temporal evolution of the wave function in the presence of isolated defects, going beyond previous quantum mechanical implementations \cite{PhysRevB.31.2479,PhysRevB.47.14842} that have yielded only formal or approximate solutions. 

We study the canonical tight-binding model (TBM), a work-horse in quantum transport studies~\cite{PhysRevB.34.3625,PhysRevA.91.062115,PhysRevE.95.032141, PhysRevLett.120.040502,das2022quantum,Das_2022}, which describes  a quantum particle hopping between nearest-neighbour sites of a one-dimensional lattice.  Heterogeneities in the form of spatially-defective sites may either affect the hopping rate between adjacent sites or the energy of a site. We focus on the latter, and in contrast to recent spectral or numerical analysis of the non-unitary time evolution due to an imaginary on-site potential~\cite{krapivsky2014survival,PhysRevA.102.012212}, we consider unitary evolution and look at closed-form time-resolved observables. While our analytical formalism works for any number of isolated defects at arbitrary locations and of arbitrary strengths, we focus on a single defect and formally proceed analogously to the case of a classical one-dimensional chain with an isolated quenched impurity studied in Ref. ~\cite{Lindenberg-pearlstein-hemenger}.
We uncover  rich and remarkable nonlinear effects, namely non-monotonic
suppression of transport, enhanced localization at distant sites, and strong sensitivity
to the initial particle position at long times. This latter fact is indeed generically observed in continuous-time quantum walks, see Ref.~\cite{mulken2011continuous} for a review.
%%%%

The paper is organised as follows. We describe the model in Sec.~\ref{sec2:model},
followed by a discussion of the defect-free dynamics in Sec.~\ref{sec3:defectFree}. The dynamics in presence of a single defect is discussed in Sec.~\ref{sec4:Defectdynamics}. In this section, we also discuss the limiting case of infinite defect strength and the case when there are multiple onsite defects present in the system. The
paper concludes with a summary and outlook in Sec.~\ref{sec5:summary}.  Some of the technical details and additional discussions are presented in the Appendices.

\section{The model}\label{sec2:model} 
We consider the TBM lattice with $N$ sites labeled as $n=0,1,\ldots, N-1$, with unit lattice spacing and periodic boundary conditions. With a defect modeled as an on-site energy $q$ at $n=n_d$,  the Hamiltonian is        
\begin{align}
H=-\sum_{n=0}^{N-1} \gamma\left( |n +1 \rangle  \langle n | + |n \rangle  \langle n+1 | \right) -  q |n_d \rangle  \langle n_d |, \label{eq:hamiltonian}
\end{align}
where $\gamma > 0$ is the hopping strength. The Wannier states $|n\rangle$ denote the state of the particle when located on site $n$, and which form a complete set of orthonormal basis. With the particle initially on site $n_0$, its state at time $t$ is $|\psi_{n_0}(t)\rangle$, with the wave function in the Wannier basis given by $\psi(n,n_0,t)\equiv\langle n|\psi_{n_0}(t)\rangle $ and the probability to be located on site $n$ given by $P_n(t)=|\psi(n,n_0,t)|^2$, satisfying $\sum_{n=0}^{N-1}P_n(t)=1~\forall~t$. Setting $\hbar=1$, the Schr\"{o}dinger evolution reads as
\begin{align}
    &\frac{\partial \psi(n,n_0,t)}{\partial t}-i \gamma \left ( \psi (n+1,n_0,t)+\psi (n-1,n_0,t)\right) =iq\delta_{n n_d} \psi(n,n_0,t). \label{Eq:Inhomogeneous_TDSE}
\end{align}

We remark that the occupation probability $P_n(t)$ remains unchanged under $q \to -q$. To show this, 
let us write the Hamiltonian in Eq.~\eqref{eq:hamiltonian} as $H(q)=H_0+q V$, with $H_0$ denoting the first term and $qV$ denoting the second term on the right hand side of Eq.~\eqref{eq:hamiltonian}. Next, we define the unitary operator $\Gamma$ as $\Gamma= (-1)^n\sum_n|n\rangle \langle n|$, with $\Gamma |n \rangle = (-1)^n |n \rangle$,  $\Gamma^\dagger=\Gamma$ and $\Gamma^\dagger \Gamma= \mathbb{I}$. Using $\Gamma H_0 \Gamma^{\dagger}=- H_0$ and $\Gamma V \Gamma^{\dagger} = V $ yields that
\begin{align}
    \Gamma H(q) \Gamma^{\dagger} = - H_0 +q V = - H(-q),
\end{align}
which implies that the unitary operator $\Gamma$ effects a unitary transformation from $H(q)$ to $H(-q)$, up to a sign. 

Next, consider the eigenvalue equation of $H(q)$ and $H(-q)$: 
\begin{align}
&H(q) |\phi_\alpha^{(q)}\rangle =  E_\alpha^{(q)}|\phi_\alpha^{(q)}\rangle, \label{eq:H_q_eigs}\\
&H(-q) |\phi_\alpha^{(-q)}\rangle =  E_\alpha^{(-q)}|\phi_\alpha^{(-q)}\rangle. \label{eq:H_minusq_eigs}
\end{align}
The first equation yields $\Gamma H(q) \Gamma^{\dagger} \Gamma |\phi_\alpha^{(q)}\rangle =  E_\alpha^{(q)} \Gamma|\phi_\alpha^{(q)}\rangle$, i.e., $~H(-q) \left( \Gamma |\phi_\alpha^{(q)}\rangle \right) = -E_\alpha^{(q)} \left( \Gamma |\phi_\alpha^{(q)}\rangle \right)$.
Comparing with Eq.~\eqref{eq:H_minusq_eigs} gives
\begin{align}
    E_\alpha^{(-q)}=-E_\alpha^{(q)},
\end{align}
and that $|\phi_\alpha^{(-q)}\rangle = e^{i \Delta} \left(\Gamma |\phi_\alpha^{(q)}\rangle\right)$, where $\Delta$ is arbitrary and real. 

We have, by definition, that
\begin{align}
    \psi_q(n,n_0,t)&=\langle n | e^{-i H(q) t}| n_0\rangle=\sum_\alpha e^{-i E_\alpha^{(q)} t} \phi_\alpha^{(q)} (n)~[\phi_\alpha^{(q)}(n_0)]^\star, 
    \label{eq:psi-0}
\end{align}
with $\langle n |\phi_\alpha^{(q)}\rangle=[\langle\phi_\alpha^{(q)}| n\rangle]^\star=\phi_\alpha^{(q)} (n)$. On the other hand, we have 
\begin{align}
    \psi_{-q}(n,n_0,t)
    &=\sum_\alpha e^{-i E_\alpha^{(-q)} t}~~ \langle n|\phi_\alpha^{(-q)}\rangle ~~\langle\phi_\alpha^{(-q)}|n_0\rangle\nonumber\\
    &=\sum_\alpha e^{-i E_\alpha^{(-q)} t}~~ \langle n|\Gamma |\phi_\alpha^{(q)}\rangle ~~\langle\phi_\alpha^{(q)}|\Gamma^\dagger|n_0\rangle\nonumber\\
    &=\sum_\alpha e^{i E_\alpha^{(q)} t} \phi_\alpha^{(q)} (n)~[\phi_\alpha^{(q)}(n_0)]^\star~(-1)^{n+n_0},
\end{align}
where we have used $E_\alpha^{(-q)}=-E_\alpha^{(q)}$ and $\langle n|\Gamma|\phi_\alpha^{(q)}\rangle = (-1)^{n}\phi_\alpha^{(q)} (n)$. Our desired result now follows straightforwardly: $P_n^{(q)}(t)=|\psi_q(n,n_0,t)|^2=|\psi_{-q}(n,n_0,t)|^2=P_n^{(-q)}(t)$.

\section{Defect-free dynamics}\label{sec3:defectFree}
The homogeneous ($q=0$) part of Eq.~\eqref{Eq:Inhomogeneous_TDSE} has the solution given by the Green's function~\cite{das2022quantum}
\begin{align}
  G(n,n_0,t)=\frac{1}{N}\sum_{k=0}^{N-1} e^{2i\gamma t \cos(2 k \pi /N)+2i\pi k (n-n_0)/N}.\label{Eq:homogeneous_sol}
\end{align}
While the probability $P_n(t)=|G(n,n_0,t)|^2$ for a fixed $n$ oscillates as a function of time, its time average reaches a steady state at long times. Denoting by $\overline{P}_n$ such a value, we have for even $N$ that $\overline{P}_n=(1/N)+(1/N^2)\left(\sum_{k=1;k \ne N/2}^{N-1}e^{2i \pi (2k-N)(n-n_0)/N}\right)$, which when evaluated yields 
\begin{align}
\overline{P}_n= \left(\frac{2}{N}-\frac{2}{N^2}\right)(\delta_{n,n_0}+\delta_{n,n_0+N/2})+\left(\frac{1}{N}-\frac{2}{N^2}\right)\left(1-\delta_{n,n_0}-\delta_{n,n_0+N/2}\right);~~\mathrm{even~}N.
\label{eq:Pn-evenN}
\end{align}
By contrast, we have 
\begin{align}
\overline{P}_n= \left(\frac{2}{N}-\frac{1}{N^2}\right)~\delta_{n,n_0}+\left(\frac{1}{N}-\frac{1}{N^2}\right)\left(1-\delta_{n,n_0}\right);~~\mathrm{odd~}N.
\label{eq:Pn-oddN}
\end{align}
Thus, $\overline{P}_n$ is homogeneous with respect to $n$, except at the initial site for both odd and even $N$ and at a site shifted by half the lattice size for only even $N$. 

In passing, let us now discuss the case $N\to \infty$. Defining $\bar k \equiv 2\pi k/N$, we note that as $N\to \infty$, the variable $\bar k$ becomes continuous, and one may convert Eq.~\eqref{Eq:homogeneous_sol} into an integral, as follows:
\begin{align}
  G(n,n_0,t)=\frac{1}{2\pi}\int_0^{2\pi}d\bar k\,e^{2i\gamma t \cos\bar k+i\bar k (n-n_0)},\label{Eq:homogeneous_sol-c}
\end{align}
which leads to the result that $P_n(t)=J_{n-n_0}^2(2\gamma t)$, where $J_n(x)$ is the Bessel function of the first kind~\cite{Dunlap:1986,Perets2008QuantumWalks}. Here, we have used the identity
$\exp(-i \Delta t \cos k) = \sum_n \exp(-i n\pi /2 + i nk) J_n(\Delta t)$~\cite{Dunlap:1986}.

The above result may also be arrived at by starting from the equivalent of Eq.~\eqref{eq:psi-0} for the defect-free case~\cite{Ni2023FirstDetection}:
\begin{align}
    G(n,n_0,t)=\sum_\alpha e^{-i E_\alpha t} \phi_\alpha (n)~[\phi_\alpha(n_0)]^\star. 
    \label{eq:psi-01}
\end{align}
To proceed, we must know the eigenvalues $E_\alpha$ and the corresponding eigenvectors $ |\phi_\alpha\rangle$ of the defect-free Hamiltonian $H_0$. In the limit $N\to \infty$, these are respectively given by $E_\alpha=-2\gamma \cos \alpha$ and  $ |\phi_\alpha\rangle=(1/\sqrt{2\pi})\sum_j e^{i\alpha j}|j\rangle$, with $\alpha$ being continuous and lying in the range $[0,2\pi)$~\cite{Das_2022,Ni2023FirstDetectionQuantumWalk}. When used in Eq.~\eqref{eq:psi-01}, we get
\begin{align}
    G(n,n_0,t)=\frac{1}{2\pi}\int_0^{2\pi}d\alpha~e^{2i \gamma t\cos \alpha+i\alpha(n-n_0)}, 
    \label{eq:psi-02}
\end{align}
which correctly reproduces Eq.~\eqref{Eq:homogeneous_sol-c}. Our result~\eqref{Eq:homogeneous_sol} goes beyond these $N\to \infty$ results, by providing the modification brought about by having finite $N$.

To quantify further the steady state, let us define the $p$-th moment of $P_n(t)$ by
$\Delta_p(t)\equiv\langle [n-n_0]^p\rangle(t)=\sum_{n=0}^{N-1}[n-n_0]^p\,P_n(t);~p>0$. Here, $p=1$ and $p=2$ correspond respectively to the mean displacement and the mean–square displacement (MSD) about the initial location~\cite{note}. Here
$[n-n_0]\equiv\min\{|n-n_0|,\,N-|n-n_0|\}$ is the minimum distance between sites $n$ and $n_0$. In terms of $\chi(n,n_0)\equiv 2\pi (k_1-k_2) (n-n_0)/N$ and $\beta\equiv 4 \gamma \sin{\delta_1}\sin{\delta_2};~\delta_1 \equiv  \pi (k_1+k_2) /N$ and $\delta_2 \equiv  \pi (k_1-k_2) /N$, we obtain from Eq.~\eqref{Eq:homogeneous_sol} that 
\begin{align}
\Delta_p(t)&=\hspace{-0.35cm}\sum_{n,k_1,k_2=0}^{N-1}\hspace{-0.4cm} \frac{[n-n_0]^p}{N^2}\cos\left[{\beta(k_1,k_2) t}\right] \cos[\chi(n,n_0)].
\label{Eq:msd1_time_defectfree_PBC}
\end{align}

Both the mean displacement and the MSD exhibit the same qualitative behaviour (see \ref{sec:AppA}, \ref{sec:AppB}); we focus here on the MSD, the standard diagnostic for tight-binding wave-packet spreading~\cite{PhysRevE.59.5214,PhysRevE.64.012301,PhysRevB.75.205120,PhysRevA.109.042213}. Considering  $\Delta_2(t)$, for small $t \le t^\star$, expanding up to terms of $O(t^2)$, we get $\Delta_2(t)=Dt^2;~D=2\gamma^2$, implying a ballistic growth. The proof is as follows.

Using translation symmetry and $\sum_{k=0}^{N-1}\cos (2\pi k(n-n_0)/N)=\delta_{n n_0}$, together with Eq.~\eqref{Eq:msd1_time_defectfree_PBC} and the small-time expansion $\Delta_2(t)=Dt^2 + \mathcal{O}(t^4)$, we get  
\begin{align}
    &D=-\frac{8\gamma^2}{N^2} \sum_{n=0}^{N-1} [n]^2\mathcal{F}(n);\label{eq:full_exp_D} \\
    &\mathcal{F}(n)\!\!\equiv\!\!\!\!\!\!\sum_{k_1,k_2=0}^{N-1}\!\!\!\! 
    \cos\!\left(\!\tfrac{2\pi n(k_1-k_2)}{N}\!\right)
    \!\sin^2\!\left(\!\tfrac{\pi (k_1+k_2)}{N}\!\right)\!\sin^2\!\left(\!\tfrac{\pi (k_1-k_2)}{N}\!\right). \label{eq:F(n)}
\end{align}
Setting $r\equiv 2\pi/N$ in Eq.~\eqref{eq:F(n)} and defining $h(n,k_1,k_2)=\cos\left[r n(k_1-k_2)\right]$ and $g^{\pm}(k_1,k_2)=\cos\left[r(k_1\pm k_2)\right]$ further simplifies Eq.~\eqref{eq:F(n)} into
\begin{align}
    \mathcal{F}(n)
    &=\frac{1}{4}\left[\mathcal{F}_1 (n)+\mathcal{F}_2(n)-\mathcal{F}^{+}_3(n)-\mathcal{F}^{-}_3(n)\right];\label{eq:full_curly_F}
\end{align}
$\mathcal{F}_1(n)\equiv \sum_{k_1,k_2=0}^{N-1} h(n,k_1,k_2)g^{+}(k_1,k_2)g^{-}(k_1,k_2)$, $\mathcal{F}_2(n)\equiv \sum_{k_1,k_2=0}^{N-1} h(n,k_1,k_2)$, and $\mathcal{F}^{\pm}_3 (n) \equiv \sum_{k_1,k_2=0}^{N-1}h(n,k_1,k_2)g^{\pm}(k_1,k_2)$. With $q \equiv k_1-k_2$, we have
\begin{align}
    \!\!\mathcal{F}^{+}_3(n)=\!\!\sum_{q=-N+1}^{N-1}\!\!\!\! \cos\left[r n q\right] \sum_{k_2=0}^{N-1}\cos\left[r (2k_2+q)\right]=0.
\end{align}
Similarly, one has $\mathcal{F}_1(n)=0$. On the other hand, we have
$\mathcal{F}_2 (n)=\left(\sum_{k_1=0}^{N-1} \cos (r n k_1) \right)^2 + \left(\sum_{k_1=0}^{N-1} \sin (r n k_1) \right)^2=N^2\delta_{n,0}$.
Also, we have
\begin{align}
\mathcal{F}^{-}_3 (n)
    &=\!\frac{1}{2} \sum_{q=-N+1}^{N-1}\!\Big\{\cos[r(n+1)q]\!+\!\cos[r(n-1)q]\Big\}\nonumber\\
    &=\frac{N^2}{2}\left(\delta_{n,N-1}+\delta_{n,1}  \right).
\end{align}
Equation~\eqref{eq:full_curly_F} together with Eq.~\eqref{eq:full_exp_D} now gives
 \begin{align}
    \!\!D=\!-\frac{8\gamma^2}{N^2} \sum_{n=0}^{N-1} [n]^2\frac{N^2}{4} \left(\!\delta_{n,0}- \frac{1}{2} \left(\delta_{n,N-1}+\delta_{n,1} \right)\right).
\end{align}
Evaluating the summation straightforwardly using the definition of $[n]$ yields $D=2 \gamma^2$ for both even and odd $N$. 

\begin{figure}
\centering
\includegraphics[scale=0.30]{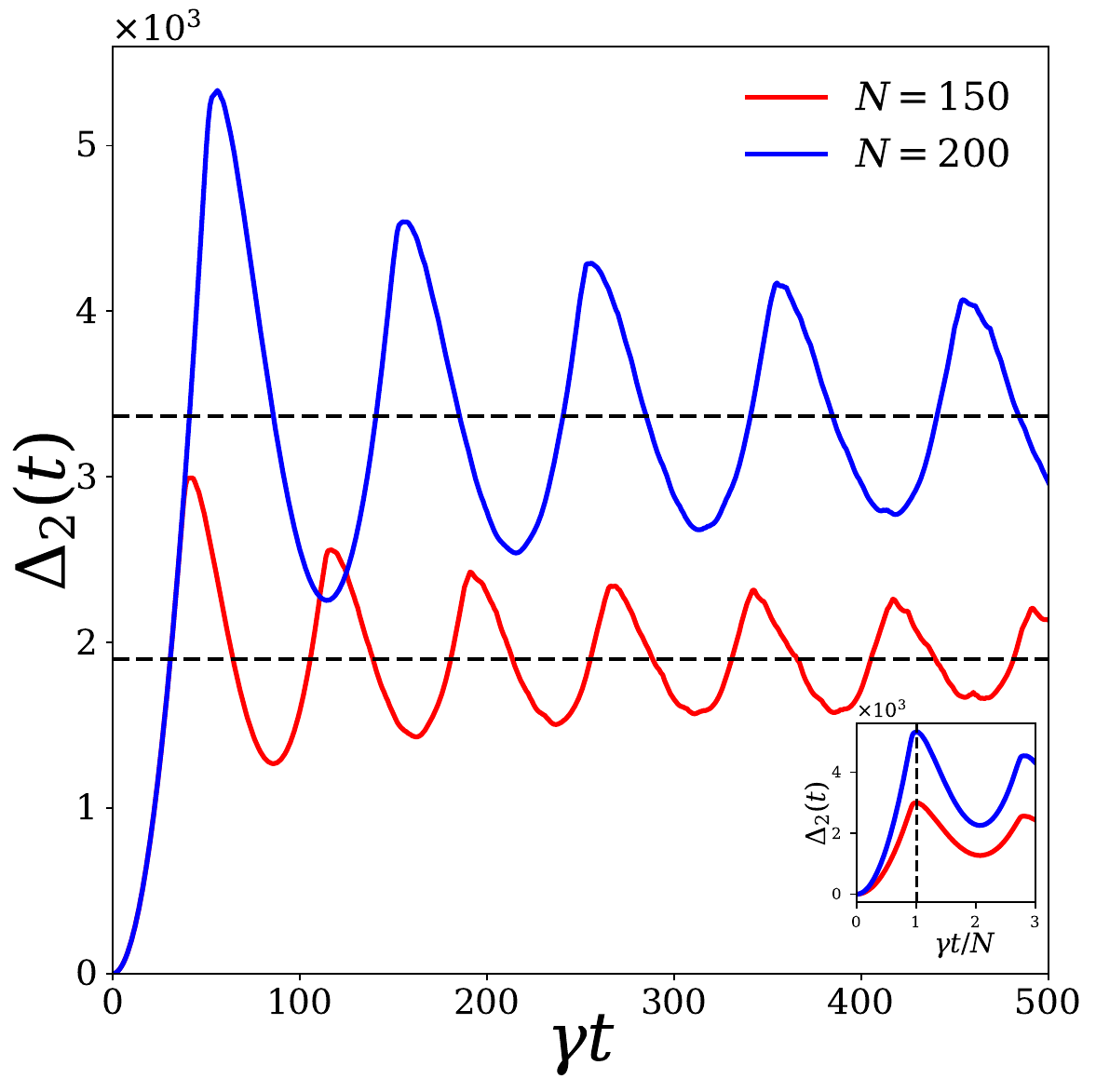}
\caption{MSD $\Delta_2(t)$ versus $t$ without defect; $\gamma=1$, and for $N=150$ and $200$, and initial site $n_0=75$ and $100$, respectively. The dashed line denotes the steady-state value $\overline{\Delta}_2$, Eq.~\eqref{steadystateMSD_defectfree_evenN}. For the timescale $t^\star$ over which oscillations appear, the inset shows $t^\star \sim N/\gamma$.}
\label{fig:Fig1}
\end{figure}

Over the timescale $t^\star$, one observes deviation from the $t^2$-dependence, resulting in oscillations, which is due to the dynamics taking place on a periodic lattice. The inset of Fig.~\ref{fig:Fig1} reveals a linear scaling $t^\star=aN/\gamma$, where $a$ is a dimensionless constant. This dependence can be rationalized through the Lieb-Robinson (LR) bound~\cite{LiebRobinson1972}, which constrains the speed of information propagation in non-relativistic quantum systems. In the TBM set-up, this bound depends critically on the range of allowed hopping of the particle between the sites~\cite{rbtb-8d27,PhysRevX.10.031009}. For the nearest-neighbour hopping of Eq.~\eqref{eq:hamiltonian}, the LR velocity is a constant, leading to light-cone-like propagation and a linear scaling of the timescale of propagation with distance. Consequently, the time $t^\star$, marking the timescale over which the dynamics becomes sensitive to the periodic boundary conditions, scales linearly with the periodic extent $N$ of the lattice. Note that the $N\to \infty$ result in Eq.~\eqref{Eq:homogeneous_sol-c} implies $\Delta_2(t)=Dt^2$ for all $t$~\cite{Dunlap:1986}. This is consistent with the result quoted above for finite $N$ (that $\Delta_2(t)=Dt^2$ for $t<t^\star$), since as $N\to \infty$, the timescale $t^\star \sim N$ diverges, and it is only the $t^2$-dependence that is observed for all times.

We now evaluate the steady-state values for the mean displacement and the MSD.
Using Eqs.~\eqref{eq:Pn-evenN} and~\eqref{eq:Pn-oddN}, we get for even $N$
\begin{align}
    \overline{\Delta}_1 &= \!\left(\frac{2}{N}-\frac{2}{N^2}\right)\! \frac{N}{2}\!+\!\left(\frac{1}{N}-\frac{2}{N^2}\right)\!\left(\sum_{n=0}^{N-1} [n-n_0] - \frac{N}{2}\right)\nonumber\\
    &=\frac{1}{2}+\frac{1}{N^2} (N-2) \sum_{n=0}^{N-1} [n-n_0]=\frac{N}{4},\label{eq:mean_steady_evenN}
\end{align}
where $\sum_{n=0}^{N-1} [n-n_0]=N^2/4$. In the same manner, we get
\begin{align}
    \overline{\Delta}_2 &=\frac{N^2+2}{12} + \frac{N}{12} -\frac{1}{3N},
\end{align}
using $\sum_{n=0}^{N-1} [n-n_0]^2=N(N^2+2)/12$. Next, for odd $N$, proceeding similarly as for even $N$, we obtain
\begin{align}
  \overline{\Delta}_1=\frac{(N-1)^2(N+1)}{4N^2}~;~\overline{\Delta}_2=\frac{(N-1)^2(N+1)}{12N},\label{steadystateMSD_defectfree_evenN}
\end{align}
where we have used $\sum_{n=0}^{N-1} [n-n_0]= (N^2-1)/4$ and $\sum_{n=0}^{N-1} [n-n_0]^2=N (N^2-1)/12$. 

\section{Dynamics in the presence of the defect}\label{sec4:Defectdynamics}
Here, we employ the defect technique, which provides a compact route to quantities that are otherwise inaccessible in classical random-walk problems. In such problems, whenever a localized perturbation is introduced, i.e., by modifying the transition probability $\mathcal{T}$ at a single site $d$, as  $\mathcal{T}^{\prime} = \mathcal{T} + \Pi $, with $\Pi$ non-zero only at $d$, the corresponding Green's function, $G^{\prime} = (I -\mathcal{T}^{\prime})^{-1}$, is related to the unperturbed one through the resolvent identity $G^{\prime} = G - G\,\Pi\,G^{\prime}$, with $G = (I - \mathcal{T})^{-1}$. Solving this single-site self-consistent equation \textit{at the defect site} yields the sought-after contribution,  $G^{\prime}_d=G_d\big[1+G_d\,\Pi\big]^{{-1}}$. By inserting it back in the resolvent identity, one captures the full response of the walk to the perturbation~\cite{kenkre2021memory}. In other words, the defect technique isolates the local modification and renders the problem analytically tractable in closed form. Such a procedure can be followed for an arbitrary (finite) number of defective sites and irrespective of the type of defect.

To apply the defect technique to the case at hand, we start with Eq.~\eqref{Eq:Inhomogeneous_TDSE}, which yields in the Laplace domain $\epsilon$, with $\widetilde{f}(\epsilon)=\mathcal{L}\left\{f(t)\right\}=\int_0^{\infty}dt~e^{-\epsilon t}f(t)$,
\begin{align}
\widetilde{\psi}(n,n_0,\epsilon) = \widetilde{G}(n,n_0,\epsilon)+iq \widetilde{G}(n,n_d,\epsilon)\widetilde{\psi}(n_d,n_0,\epsilon).
\label{eq:defect-eqn}
\end{align}
To proceed further, we require to know $\widetilde{\psi}(n_d,n_0,\epsilon)$, which in turn depends on $\widetilde{\psi}(n,n_0,\epsilon)$. The defect technique allows to bypass this apparent circular dependence by setting $n=n_d$ in Eq.~\eqref{eq:defect-eqn}, giving $\widetilde{\psi}(n_d,n_0,\epsilon)=\widetilde{G}(n_d,n_0,\epsilon)/\left(1 - i q \widetilde{G}(n_d,n_d,\epsilon)\right)$. Equation~\eqref{eq:defect-eqn} then yields
    \begin{align}
\!\!\!\!\!\widetilde{\psi}(n,n_0,\epsilon) \!=\! \widetilde{G}(n,n_0,\epsilon)\!+\! \widetilde{G}(n,n_d,\epsilon)\frac{i q \widetilde{G}(n_d,n_0,\epsilon)}{1- i q \widetilde{G}(n_d,n_d,\epsilon)}.\label{eq:solution_psi_defect}
\end{align}
 By defining the function $\widetilde{\Phi}(\epsilon)\equiv i q \widetilde{G}(n_d,n_0,\epsilon)/(1- i q \widetilde{G}(n_d,n_d,\epsilon))$ and noting that~\cite{PhysRevX.10.021045} 
\begin{align}
\!\!\!\widetilde{G}(n,n_0,\epsilon)\!=\!\frac{1}{2 i \gamma}\frac{T_{N-|n-n_0|}(\epsilon/2 i \gamma)+T_{|n-n_0|}(\epsilon/2 i \gamma)}{\big[(\epsilon/2 i \gamma)^2-1\big]U_{N-1}(\epsilon/2 i \gamma)}, \label{eq:homogeneous_G_laplace}
\end{align}
where $T_m(z)$ and $U_m(z)$ are the Chebyshev polynomials of the first and the second kind, respectively, one has $
    \widetilde{\Phi}(2 i \gamma \epsilon)=[q/(2 \gamma)] \mathcal{P}(\epsilon)/\mathcal{Q}(\epsilon)$,
with $\mathcal{P}(\epsilon)\equiv T_{N-|n_d-n_0|}(\epsilon)+T_{|n_d-n_0|}(\epsilon)$ and $\mathcal{Q}(\epsilon)\equiv (\epsilon^2-1)U_{N-1}(\epsilon)- [q/(2 \gamma)] [T_{N}(\epsilon)+1]$.
Performing the Laplace inversion, we get
\begin{align}
    \!\!\Phi(t)=iq ~\hspace{-0.3cm} \sum_{x_j \in \text{poles}} ~\hspace{-0.38cm}\mathrm{Res}\left[\frac{\mathcal{P}(\epsilon)}{\mathcal{Q}(\epsilon)}~e^{2 i \gamma \epsilon t} \right]_{\epsilon=x_j}\!\!\!\!=\!iq ~\hspace{-0.3cm}\sum_{x_j\in \mathrm{poles}}\!\!\!f_j e^{2 i \gamma x_j t},\label{eq:phi_t_q_finite}
\end{align}
with $f_j\equiv\mathcal{P}(x_j)/\mathcal{Q}'(x_j)$, the prime denoting derivative, and poles $x_j$ given by the roots of $\mathcal{Q}(x)$. With $\mathcal{P}(\epsilon)$ and $\mathcal{Q}(\epsilon)$ both real, we have $\Phi^*(t)=-i\,q\sum_{x_j\in \mathrm{poles}} f_j e^{-2 i \gamma x_j t}$,  and the star denoting complex conjugation. 

\begin{figure}
       \centering
    \includegraphics[scale=0.24]{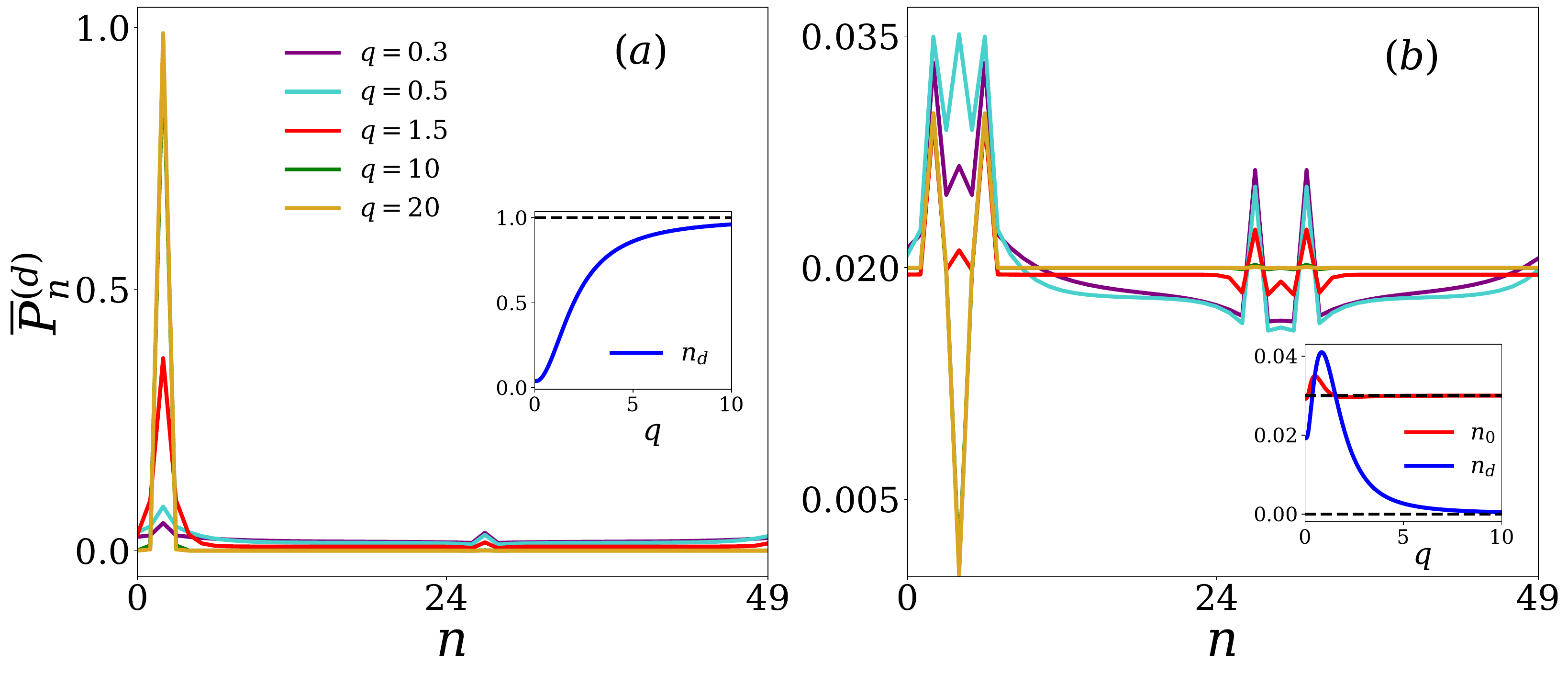}      \caption{Steady-state probability $\overline{P}_n^{(d)}$ versus $n$ for defect strength $q = 0.3,0.5,1.5,10,20$ on a lattice of size $N=50$ and $\gamma=1$. In panel (a),  the initial site coincides with the defect site, $n_d = n_0 = 2$. Inset: Probability at $n_d$ increases monotonically with $q$ and asymptotically reaches $1$ as $q \to \infty$. In panel (b), the initial site is located away from the defect site: $n_d = 4$, $n_0 = 2$. Inset: Non-monotonic dependence of the probability at sites $n_0$ and $n_d$ on the defect strength; the probability at both sites asymptotically reaches values given by Eq.~\eqref{eq:Pnd-qinf}, see the dashed lines.}
 \label{fig:Fig2}
\end{figure}
 
Inverting Eq.~\eqref{eq:solution_psi_defect} to the time domain gives $\psi(n,n_0,t)=G(n,n_0,t) + \int_0^t dt_1 G(n,n_d,t-t_1) \Phi(t_1)$. One then obtains the occupation probability in the presence of the defect as
\begin{align}
&P_n^{(d)}(t)= P_n(t) + I_n^{(d)}(t)+K_n^{(d)}(t)
\label{eq:Probability_with_defect-1};
\end{align}
$I_n^{(d)}(t)\equiv G^*(n,n_0,t) A(n,n_d,t)+G(n,n_0,t)A^*(n,n_d,t)$, $K_n^{(d)}(t)\equiv |A(n,n_d,t)|^2$, $A(n,n_d,t) \equiv \int_0^t dt_1~G(n,n_d,t-t_1)\Phi(t_1)$. Let us note that both $I_n^{(d)}(t)$ and $K_n^{(d)}(t)$ depend on $q$, that normalization of $P_n^{(d)}(t)$ demands that $\sum_n \left[I_n^{(d)}(t)+K_n^{(d)}(t)\right]=0$, and that $P_n^{(d)}(t)$ is a function of all possible separations: $|n-n_0|$, $|n-n_d|$ and $|n_0-n_d|$.
 Equation~\eqref{eq:Probability_with_defect-1} is our exact expression for the site-occupation probability in presence of the defect, yielding directly observable quantities such as the mean displacement and the MSD. 

 Equation~\eqref{eq:Probability_with_defect-1}, which constitutes one of our main results, quantifies how the introduction of a single onsite defect fundamentally alters the occupation probability, through a \textit{nonlinear} modification of the defect-free case. Indeed, from the structure of $\Phi(t)$ in Eq.~\eqref{eq:phi_t_q_finite}, it is evident that $I_n^{(d)}(t)$ and $K_n^{(d)}(t)$ are nonlinear functions of $q$. Hence, the quantity $P_n^{(d)}(t)-P_n(t)$ is also a nonlinear function of $q$. In the steady state, one has $\overline{P}_n^{(d)}=\overline{P_n} + \overline{I}_n^{(d)}+\overline{K}_n^{(d)}$, where the quantities $\overline{I}_n^{(d)}$ and $\overline{K}_n^{(d)}$ can be computed exactly using the expressions for $I_n^{(d)}(t)$ and $K_n^{(d)}(t)$ (see~\ref{sec:AppC}, \ref{sec:AppD}), to obtain
\begin{align}
\overline{I}_n^{(d)}
&= \frac{q}{N^2}
\sum_{j\in\mathrm{poles}} f_j \Bigg\{
\frac{1}{\mathcal{C}_0(x_j)}+2\!\sum_{k=1}^{N-1}\! 
\frac{1}{\mathcal{C}_k(x_j)}\cos\!\left[\frac{2\pi k}{N}(n-n_d)\right]
\!\cos\!\left[\frac{2\pi k}{N}(n-n_0)\right]\!\!
\Bigg\},\label{eq:I_n_d_qfinite}
\end{align}
\vspace{-0.6cm}
\begin{align}
\overline{K}_n^{(d)}
&= \frac{q^2}{4N^2}\sum_{j\in\mathrm{poles}}\sum_{k_1,k_2=0}^{N-1}
   \frac{f_j^2}{\mathcal{C}_{k_1}(x_j)\,\mathcal{C}_{k_2}(x_j)}
   \cos\!\big[\chi(n,n_d)\big]+ \frac{q^2}{4N^2}\sum_{j,r\in\mathrm{poles}}
   \Bigg\{
     \sum_{k=0}^{N-1}\frac{f_j f_r}{\mathcal{C}_{k}(x_j)\,\mathcal{C}_{k}(x_r)}
     \nonumber\\
&+\sum_{k=1}^{N-1}\frac{f_j f_r}{\mathcal{C}_{k}(x_j)\,\mathcal{C}_{k}(x_r)}
     \cos\left[\frac{4\pi k (n-n_d)}{N}\right]
   \Bigg\},\label{eq:K_n_d_qfinite}
\end{align}
with $\mathcal{C}_{k_{\alpha}}(x_j) \equiv \gamma [\cos({2 \pi k_{\alpha} /N}) -x_j ]$ and $\chi(n,n_d)\equiv 2\pi (k_1-k_2) (n-n_d)/N$, and $x_j$ being the roots of $\mathcal{Q}(x)$, as defined after Eq.~\eqref{eq:phi_t_q_finite}.
\begin{figure}
       \centering
       \includegraphics[scale=0.24]{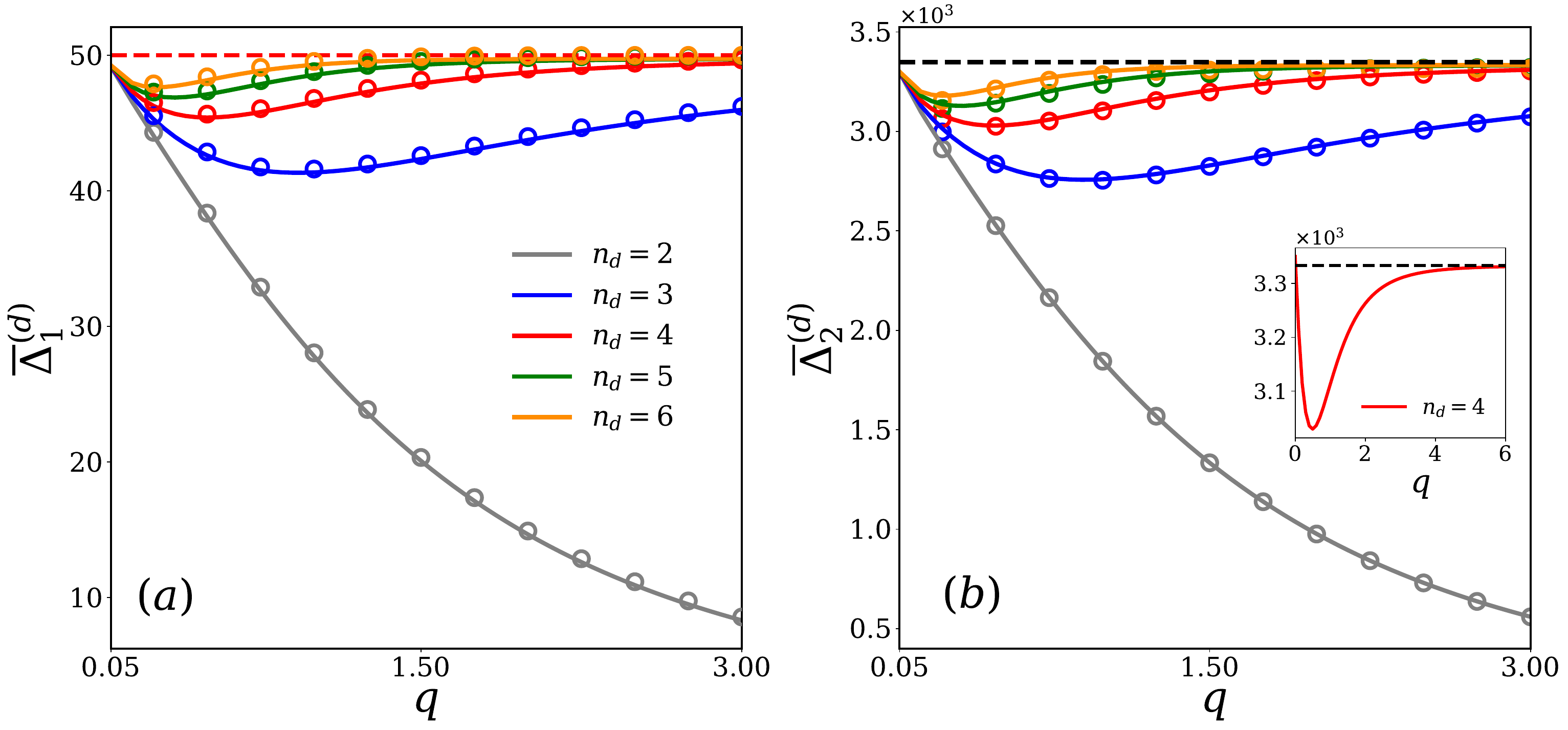}
        \caption{Steady-state mean displacement $\overline{\Delta}^{(d)}_1$ (panel (a)) and MSD $\overline{\Delta}^{(d)}_2$ (panel (b)) versus $q$ with varying defect location $n_d=2,3,4,5,6$, initial location $n_0=2$, system size $N=200$, $\gamma=1$. The lines correspond to analytical results obtained using the steady-state version of Eqs.~\eqref{eq:average-Mean-q} (panel (a)) and \eqref{eq:average-MSD-q} (panel(b)), while points denote numerical results from unitary evolution of the dynamics. In the inset of panel (b), for the case of $n_d=4$, we show $\overline{\Delta}^{(d)}_2$ asymptotically approaching the $q\to\infty$ value (dashed line).} \label{fig:Fig3}
\end{figure}

To highlight the salient features of $\overline{P}_n^{(d)}$, first consider the case $n_0=n_d$. Here, $\overline{P}_n^{(d)}$ is a function solely of $|n-n_0|$ and is peaked at $n_d$; in fact, Fig.~\ref{fig:Fig2}(a) shows that $\overline{P}_{n_d}^{(d)}$ grows with the increase of $q$ (with a concomitant decrease of probability at all other sites, owing to normalization). This localization of the particle at the defect site, which is enhanced on increase of $q$, is trivially the result of having been placed at the defect site. A completely contrasting and richer picture is offered by the choice $n_0 \ne n_d$. In this case, strikingly, $\overline{P}_{n}^{(d)}~\forall~n$ is a non-monotonic function of increasing $q$, see Fig.~\ref{fig:Fig2}(b), implying in particular that defect-induced localization at $n_d$ does not grow monotonically with defect strength, see inset of Fig.~\ref{fig:Fig2}(b); the $q\to \infty$ limit is discussed below. In other words, the initial position of the particle leaves an enduring and unmistakable imprint on the dynamics that never dies out, even at long times. Note that localization in our system is a dynamical trapping-like phenomenon; what it implies is that in the limit of long times, the system spends a disproportionate amount of time near certain sites because transition away from them is suppressed.

To elucidate further the effects of the defect on the steady state, we consider the mean displacement and the MSD, given respectively on using Eq.~\eqref{eq:Probability_with_defect-1} by
 \begin{align}
&\Delta^{(d)}_1(t)=\Delta_1(t)+\mathcal{I}^{(1)}(t)+\mathcal{K}^{(1)}(t),\label{eq:average-Mean-q}\\
     &\Delta^{(d)}_2(t)=\Delta_2(t) + \mathcal{I}^{(2)}(t)+\mathcal{K}^{(2)}(t);
     \label{eq:average-MSD-q}
 \end{align}
 $(\mathcal{I}^{(p)}(t),\mathcal{K}^{(p)}(t))\equiv\sum_{n=0}^{N-1} [n-n_0]^p(I_n^{(d)}(t),K_n^{(d)}(t))$. For $n_0\ne n_d$, the steady-state values $\overline{\Delta}^{(d)}_1$ and $\overline{\Delta}^{(d)}_2$ vary non-monotonically with $q$, exhibiting a minimum as shown in Fig.~\ref{fig:Fig3}; similar to the case without defect, the mean and the MSD show qualitatively similar behavior. By contrast, for $n_0=n_d$, both decrease monotonically with increasing $q$ as a result of enhanced localization. Next, we discuss two salient aspects of the dynamics involving defects in the following subsections.

 Let us discuss the result in Eq.~\eqref{eq:average-MSD-q}, and view it in the light of the corresponding defect-free answer $\Delta_2(t)$ plotted in Fig.~\ref{fig:Fig1}. From the form of the functions $I_n^{(d)}(t)$ and $K_n^{(d)}(t)$ (see the text following Eq.~\eqref{eq:Probability_with_defect-1}), and hence of the functions $\mathcal{I}^{(2)}(t)$ and $\mathcal{K}^{(2)}(t)$, it is evident that in presence of the defect, the qualitative behavior changes substantially already for small $q$, since the functions $I_n^{(d)}(t)$ and $K_n^{(d)}(t)$ become non-zero and in a nonlinear manner as soon as $q\ne 0$.
 
\subsection{$q \to \infty$ limit}\label{subsec:q_infinity}

We now focus on the $q\to \infty$ limit of the dynamics. In this limit, for $n_0=n_d$, we get complete localization at the defect site,  $\overline{P}_n^{(d)}=\delta_{n,n_d}$. On the other hand, for $n_0 \ne n_d$, Eq.~\eqref{eq:solution_psi_defect} reads as 
\begin{align}
\!\!\!\!\!\widetilde{\psi}(n,n_0,\epsilon) \!=\! \widetilde{G}(n,n_0,\epsilon)\!-\! \widetilde{G}(n,n_d,\epsilon)\frac{ \widetilde{G}(n_d,n_0,\epsilon)}{\widetilde{G}(n_d,n_d,\epsilon)}.\label{eq:solution_psi_defect-1}
\end{align}
The propagator ratio on the right hand side of Eq.~\eqref{eq:solution_psi_defect-1}, with $f_N(x)\equiv T_{N}(x)+1$, gives 
\begin{align}
\frac{ \widetilde{G}(n_d,n_0,\epsilon)}{ \widetilde{G}(n_d,n_d,\epsilon)}=\frac{T_{N-|n_d-n_0|}(\epsilon/2i\gamma)+T_{|n_d-n_0|}(\epsilon/2i\gamma)}{f_N(\epsilon/2i\gamma)};\label{eq:laplace_infq1}
\end{align}
$f_N(x)=T^{2}_{N/2}(x)$ for even $N$ and $f_N(x)=(1+x) V^2_{(N-1)/2}(x)$ for odd $N$, $V_{m}(x)$ being the Chebyshev polynomial of the third kind. Laplace inversion of Eq.~\eqref{eq:laplace_infq1} to the time domain leads to the integral $1/(2\pi i)\oint d\epsilon \big[T_{N-|n_d-n_0|}(\epsilon/2i\gamma)+T_{|n_d-n_0|}(\epsilon/2i\gamma)\big]e^{\epsilon t}/f_{N}(\epsilon/2i\gamma)$. Knowing the roots $\mathrm{x}_k$ of $f_N(x)$, we evaluate the integral by using the residue theorem. For even $N$, using $f_N(x)=T^{2}_{N/2}(x)$ implies a pole of order $2$ at $x=\mathrm{x}_k$. Denoting the numerator of Eq.~\eqref{eq:laplace_infq1} as $g(x)$, we then have
\begin{align}
&\mbox{Res}\left(\frac{g(x)}{f_N(x)}\right)\bigg|_{x=\mathrm{x}_k}\!\!\!\!=
\frac{6\,g^{\prime}(x)f_N^{\prime\prime}(x)-2\,g(x)f_N^{\prime\prime\prime}(x)}{3\big[f^{\prime\prime}(x)\big]^2}\bigg|_{x=\mathrm{x}_k}; \\
&f_N^{\prime\prime}(x=\mathrm{x}_k)=N^2\left\{2 \sin^2\left[\frac{(2k-1)\pi}{N}\right]\right\}^{-1},\nonumber\\
    &g^{\prime}(x=\mathrm{x}_k)=N\frac{\sin\left[|n_d-n_0|\frac{(2k-1)\pi}{N}\right]}{\sin\left[\frac{(2k-1)\pi}{N}\right]}e^{2 i \gamma \cos\left[\frac{(2k-1)\pi}{N}\right]t},\nonumber 
\end{align}
and $g(x=\mathrm{x}_k)=0$. For odd $N$, using $f_N(x)=(1+x)V^{2}_{N/2}(x)$ implies a pole of order $2$ at $x=\mathrm{x}_k$. Note that $x=-1$ makes $g(x)/f_N(x)$ regular since $g(x=-1)=0$. For odd $N$, one obtains the same expressions for $f_N^{''}(x)$ and $g'(x)$ as for even $N$. We then obtain for both even and odd $N$ the function $\Phi(t)=\mathcal{L}^{-1}\left\{-\widetilde{G}(n_d,n_0,\epsilon)/\widetilde{G}(n_d,n_d,\epsilon)\right\}$ in the limit $q\to \infty$ as ($\theta_k \equiv \pi(2k-1)/N$) 
\begin{align}
\Phi(t)&=-i \tilde{\gamma} \sum_{k=1}^{\lfloor N/2 \rfloor} F_k e^{2i\gamma t \cos(\theta_k)}, \label{eq:Phi_t_qinf_maintext1}
\end{align}
with $F_k\equiv\sin(|n_d-n_0| \theta_k )\sin(\theta_k)$ and $\tilde{\gamma}\equiv 4\gamma/N$, and $\lfloor \cdot \rfloor$ denotes the greatest integer function, mapping a real number to the greatest integer lesser than it or itself. Note that $\Phi(t)$ has the same form as Eq.~\eqref{eq:phi_t_q_finite}. Following the same steps that led from Eq.~\eqref{eq:phi_t_q_finite} to Eq.~\eqref{eq:Probability_with_defect-1}, we obtain the $q\to \infty$ limit of Eq.~\eqref{eq:Probability_with_defect-1}. We denote the $q\to \infty$ counterparts of $\overline{I}_n^{(d)}$ and $\overline{K}_n^{(d)}$ as $\overline{\mathbb{I}}_n^{(d)}$ and $\overline{\mathbb{K}}_n^{(d)}$ respectively. Taking further the time average, one obtains the following exact results (see \ref{sec:AppE})
\begin{align}
&\overline{\mathbb{I}}_n^{(d)}=\begin{cases}-\frac{N-2}{N^2}-\frac{1}{N} \left(\delta_{n,n_0}+\delta_{n,n_d}+\delta_{n,n_0+\frac{N}{2}}+\delta_{n,n_d+\frac{N}{2}}\right);~~\mathrm{even~}N,\\
-\frac{N-2}{N^2}-\frac{1}{N}\left( \delta_{n,n_0}+\delta_{n,n_d}\right);~~\mathrm{odd~}N,\label{eq:I_n_d_qinfinite_even_odd_N}
\end{cases}\\
&\overline{\mathbb{K}}_n^{(d)}=\frac{4 {\gamma}^2}{N^4} \Biggl[\sum_{l=1}^{\lfloor N/2 \rfloor}F_l^2~\left|\sum_{k=0}^{N-1} \frac{e^{2i\pi k(n-n_d)/N}}{\mathbb{C}_{k}(\mathrm{x}_l)}  \right|^2+\sum_{k=0}^{N-1} R_{k}^2+\sum_{k=1}^{N-1} R_{k}^2 \cos{(4 \pi k (n-n_d)/N)}\Biggr],\label{eq:K_n_d_qinfinite_even_odd_N}
\end{align}
with $\mathbb{C}_{k_{\alpha}}(\mathrm{x}_l) \equiv \gamma [\cos({2 \pi k_{\alpha} /N}) -\mathrm{x}_l ]$, $\mathrm{x}_l=\cos[(2l-1)\pi/N]$ where $l=1,2,..., \lfloor N/2 \rfloor$ and the quantity $R_k$ is defined in \ref{sec:AppE} and $F_l$ is defined after Eq.~\eqref{eq:Phi_t_qinf_maintext1}. Using the defect-free steady state site-occupation probability $\overline{P}_n$ and $\overline{P}_n^{(d)}=\overline{P_n} + \overline{\mathbb{I}}_n^{(d)}+\overline{\mathbb{K}}_n^{(d)}$, we  obtain $\overline{P}_n^{(d)}$ in the limit $q\to \infty$ as
\begin{align}
\overline{P}_n^{(d)}&=\frac{3}{2 N}\left(\delta_{n,2n_d-n_0}+\delta_{n,n_0}\right)+\frac{1}{N}\left(1-\delta_{n,2n_d-n_0}-\delta_{n,n_0}-\delta_{n,n_d} \right).
    \label{eq:Pnd-qinf}
\end{align}
\begin{figure}
       \centering
        \includegraphics[scale=0.24]{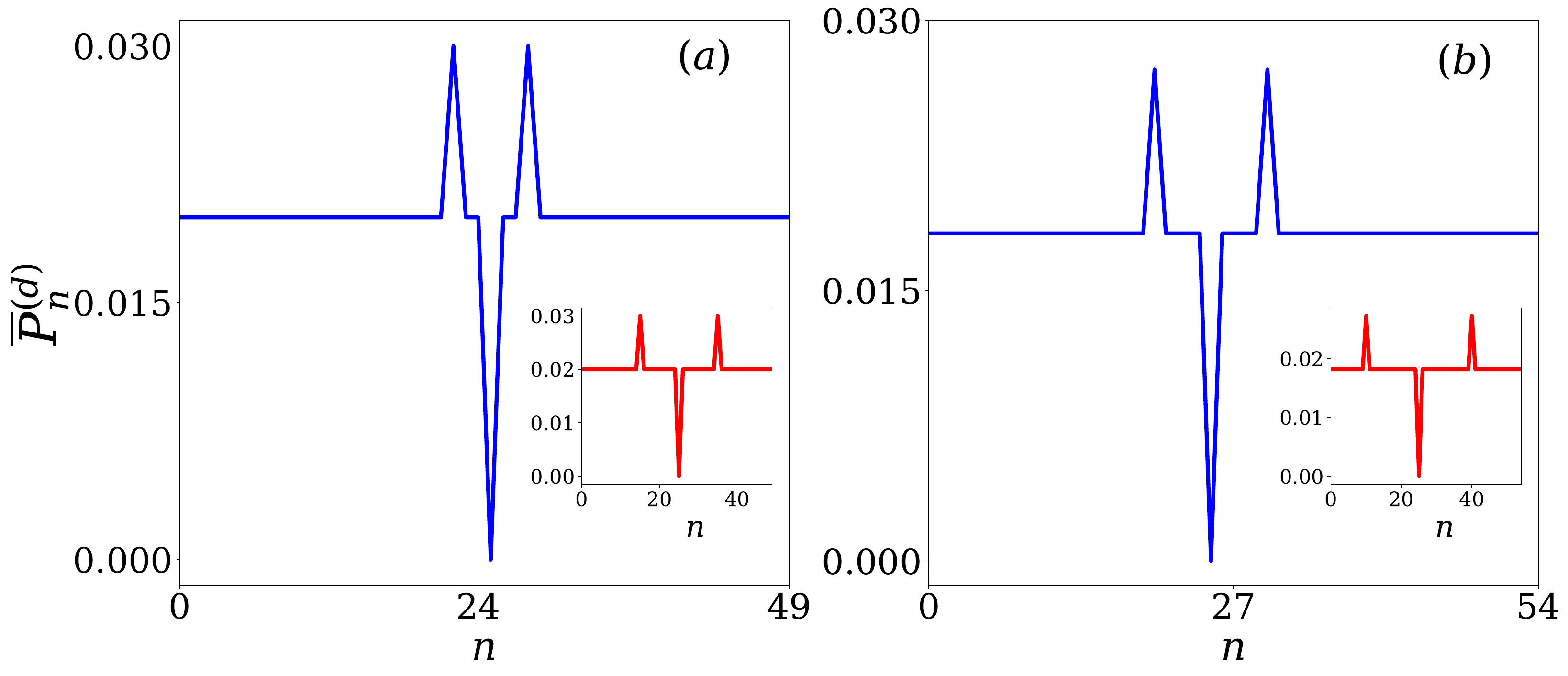}
        \caption{$\overline{P}_n^{(d)}$ versus $n$ for $q \to \infty$ and for even and odd $N$. Panel (a): $N=50$, $n_d=25$, $n_0=22$; Panel (b): $N=55$, $n_d=25$, $n_0=30$. Insets: $\overline{P}_n^{(d)}$ versus $n$ for different initial locations, $n_0=15$ (a) and $n_0=40$ (b). In all cases, $\gamma=1$.} \label{fig:App1}
\end{figure}

Equation~\eqref{eq:Pnd-qinf}, our second main result, implies that in the limit $q\to \infty$, the particle has zero (time-averaged)  probability to be found at the defect site and a constant probability to be found at all other sites, except two special sites at distance $|n_d-n_0|$ on either side of the defect site. At these special sites, the probability is equal and invariant as one moves the initial site away from the defect site (see Fig.~\ref{fig:App1}). This counterintuitive steady-state property, in which the occupation probability is strongly suppressed at the defect site and enhanced at distant sites, is a manifestation of \textit{nonlocal} transport properties that are characteristics of quantum systems even when the Hamiltonian is strictly nearest-neighbour as uncovered in Ref.~\cite{PhysRevB.18.4064}. Equation~\eqref{eq:Pnd-qinf} implies $\overline{\Delta}^{(d)}_1=N/4$, $\overline{\Delta}^{(d)}_2=(N^2+2)/12$ for even $N$, and $\overline{\Delta}^{(d)}_1=N/4-1/(4N)$, $\overline{\Delta}^{(d)}_2=(N^2-1)/12$ for odd $N$. One may wonder whether taking the limit $q\to \infty$ at the end of the calculation, as we have done here, is equivalent to studying the model by imposing at the outset a reflecting boundary condition at the defect site. This issue is discussed in~\ref{sec:reflecting}.
 
\subsection{Multiple onsite defects}\label{subsec:multiple_defects} 
The Laplace-domain wave function in the presence of $\mathcal{N}$ defects, with $q_k$ denoting the strength of the defect on site $n_{d_k}$, reads 
\begin{equation}
\widetilde{\psi}(n,n_0,\epsilon)\! =\! \widetilde{G}(n,n_0,\epsilon) + i \sum_{k=1}^{\mathcal{N}}\! q_k \widetilde{G}(n,n_{d_k},\epsilon)\, \widetilde{\psi}(n_{d_k},n_0,\epsilon).\label{eq:wave_function_laplace_Ndefects}
\end{equation}
In analogy with Eq.~\eqref{eq:solution_psi_defect}, the above equation can be compactly expressed in a matrix form as
$\mathbf{M}\,\mathbf{\Psi} = \mathbf{G}$, with $\mathbf{M}_{ij} \equiv \delta_{ij} - i q_j\, \widetilde{G}(n_{d_i},n_{d_j},\epsilon)$, vector component $\mathbf{\Psi}_i \equiv \widetilde{\psi}(n_{d_i},n_0,\epsilon)$ and $\mathbf{G}_i \equiv \widetilde{G}(n_{d_i},n_0,\epsilon)$. One may express $\widetilde{\psi}(n_{d_k},n_0,\epsilon)$ for any defect site $n_{d_k}$ entirely in terms of combinations of the homogeneous Green’s function $\widetilde{G}$ evaluated at the defect sites and the initial site. Upon Laplace inversion, this prescription then yields a closed-form expression for the wave function in the time domain. For example, for  $\mathcal{N}=2$, with the defects at sites $n_{d_1}$ and ${n_{d_2}}$ with respective strengths $q_1$ and $q_2$, one may get the full wave function in the Laplace domain as
\begin{align}
 &\widetilde{\psi}(n,n_0,\epsilon) = \widetilde{G}(n,n_0,\epsilon)\!+\!\frac{1}{Q(x)}{ i \Big[q_1 R_{n_{d_1}}(x)\!+\!q_2 R_{n_{d_2}}(x)\Big]}\nonumber\\
 &+\frac{q_1 q_2}{Q(x)} \left[P_{n_{d_1}}(x) \widetilde{G}(n,n_{d_1},\epsilon)+P_{n_{d_2}}(x) \widetilde{G}(n,n_{d_2},\epsilon)\right], \label{eq:psi_2defects_laplace}
\end{align}
with $x\equiv\epsilon/(2 i\gamma)$, and the quantities $R_{n_{d_i}}$, $P_{n_{d_i}}$, and $Q$ are given in \ref{sec:AppF}. One may easily check that on putting $q_2=0$ (single-defect case), Eq.~\eqref{eq:psi_2defects_laplace_appendix} reduces exactly to Eq.~\eqref{eq:solution_psi_defect} of the main text, yielding the wave function in the Laplace domain in the presence of a single defect.

\section{Summary and outlook}\label{sec5:summary}
In summary, we have studied the effect of a single localized defect on the unitary dynamics of a quantum particle moving on a one-dimensional periodic lattice. The proposed framework allows for the exact quantification of some of the non-trivial effects of the dynamics, whereby a local defective energy site of the lattice affects the spatial distribution of the quantum particle away from it. When the initial particle position coincides with the defect, the spreading of the particle wave function is monotonically suppressed with increasing defect strength. Strikingly, placing the particle initially away from the defect site renders the dependence non-monotonic. We emphasize that the nonlinear initial-condition dependence of the steady-state observables reported here is inherently of quantum origin. In contrast, for a classical continuous-time random walk on a one-dimensional lattice with a symmetric, partially permeable barrier between two sites (see Sec.~\ref{sec:AppG} and Ref.~\cite{PhysRevResearch.4.L032039}), steady-state observables of the type considered here are entirely independent of the barrier permeability and initial condition. In the limit of infinite defect-strength, we have obtained analytical expressions for the occupation probability and the mean and the MSD, and have shown that when the defect site does not coincide with the initial state, one obtains enhanced occupation at remote sites, a demonstration of steady-state nonlocality in quantum transport.

Methodologically, the analytical strategy proposed here is rather flexible. While we have focused on one on-site energy defect, the formalism developed in this work extends naturally to not only multiple defect sites but also combinations of different defect types (hopping, on-site, conformal~\cite{cnfdefect2023}), thus making it amenable to a wide class of single-particle quantum-transport problems. This broad applicability to finite lattice systems makes it a powerful quantitative tool to test predictions in current cold-atom and photonic-lattice experiments. Another direction to pursue would be to study the effects of dephasing noise, a representative recent study being the case of the tight-binding model with an on-site defect of either linear or nonlinear nature, which is subject to local dephasing noise implemented as random phase kicks~\cite{Das:2026}. In passing, we note that Ref.~\cite{Segev2013AndersonLight} discusses how disorder in finite photonic waveguide lattices can halt wave transport via interference, producing Anderson localization where initially spreading wave packets remain spatially confined. These experiments highlight that even in relatively small, finite systems, disorder alone can strongly reshape dynamics, thus providing a close conceptual parallel to the present work, where similar localization-like effects arise from a single defect rather than distributed disorder.

\section{Acknowledgements}\label{sec6:acknowledge}   We acknowledge useful discussions with Nitant Kenkre, Soumya Kanti Pal and Sayan Roy, generous allocation of computational resources of the Department of Theoretical Physics, TIFR, assistance of Kapil Ghadiali and Ajay
Salve, and the financial support of the Department of Atomic Energy,
Government of India under Project Identification No. RTI 4002. LG acknowledges funding from the Natural Environment Research
Council (NERC) Grant No. NE/W00545X/1. AA thanks the visiting
PhD fellowship program, Nordita, for his visit from September to October 2025, when this paper was being finalized.
\vspace{0.5cm}
\bibliography{paper}

\begin{thebibliography}{10}
\bibitem{note}
$\Delta_p(t)$ with $p=1$ and $p=2$ correspond respectively to the first and the second moment of $P_n(t)$ with respect to the initial location, which with slight abuse of notation we refer to as the mean displacement and the mean--square displacement (MSD) about the initial location, respectively.
\end{thebibliography}
\bibliographystyle{unsrt}
\clearpage
\appendix

\section{Analysis of Eq.~(4) of the main text}
\label{sec:AppA}
Exploiting the translation symmetry of the lattice so that one may choose $n_0=0$. We can write $\Delta_p(t)$ from Eq. \eqref{Eq:msd1_time_defectfree_PBC} of the main text as a conditional sum:
\begin{align}
\Delta_p(t)&=\sum_{n=1}^{N-1}\, \frac{[n]^p}{N^2}\sum_{x=0}^{2N-2}\sum_{y=-N+1}^{N-1} \Theta_{x,y} ~~\cos\left[{4\gamma t\sin\left(\frac{\pi x}{N}\right)\sin\left(\frac{\pi y}{N}\right)}\right] \times\cos\left(\frac{2\pi y n}{N}\right),
\label{Eq:msd_resummed_new}
\end{align}
with $x\equiv k_1+k_2$ and $y\equiv k_1-k_2$ and $\Theta_{x,y}=1$ if $\frac{x+y}{2}, \frac{x-y}{2} \in \{0,\ldots,N-1\}$, and is zero otherwise. Equation~\eqref{Eq:msd_resummed_new} gives
\begin{align}
     \Delta_p(t)
     &=\!\!\frac{1}{N^2}\!\!\sum_{x=0}^{2N-2}\!\sum_{y=-N+1}^{N-1} \hspace{-0.3cm}\Theta_{x,y\neq0}\cos\!\left[{4\gamma t\sin\left(\frac{\pi x}{N}\right)\!\sin\left(\frac{\pi y}{N}\right)}\!\right]  \sum_{n=1}^{N-1}[n]^p\cos\left(\frac{2\pi y n}{N}\right)+\sum_{n=1}^{N-1}\, \frac{[n]^p}{N}\label{eq:defectfree_Delta_p_t_explicit_exp1},
\end{align}
where $x \in \{0,2,\ldots,2N-2\}$ for $y=0$ yields $\sum_{x=0}^{2N-2}\Theta_{x,y=0}=N$. 
\begin{figure}[h]
       \centering
       \includegraphics[scale=0.24]{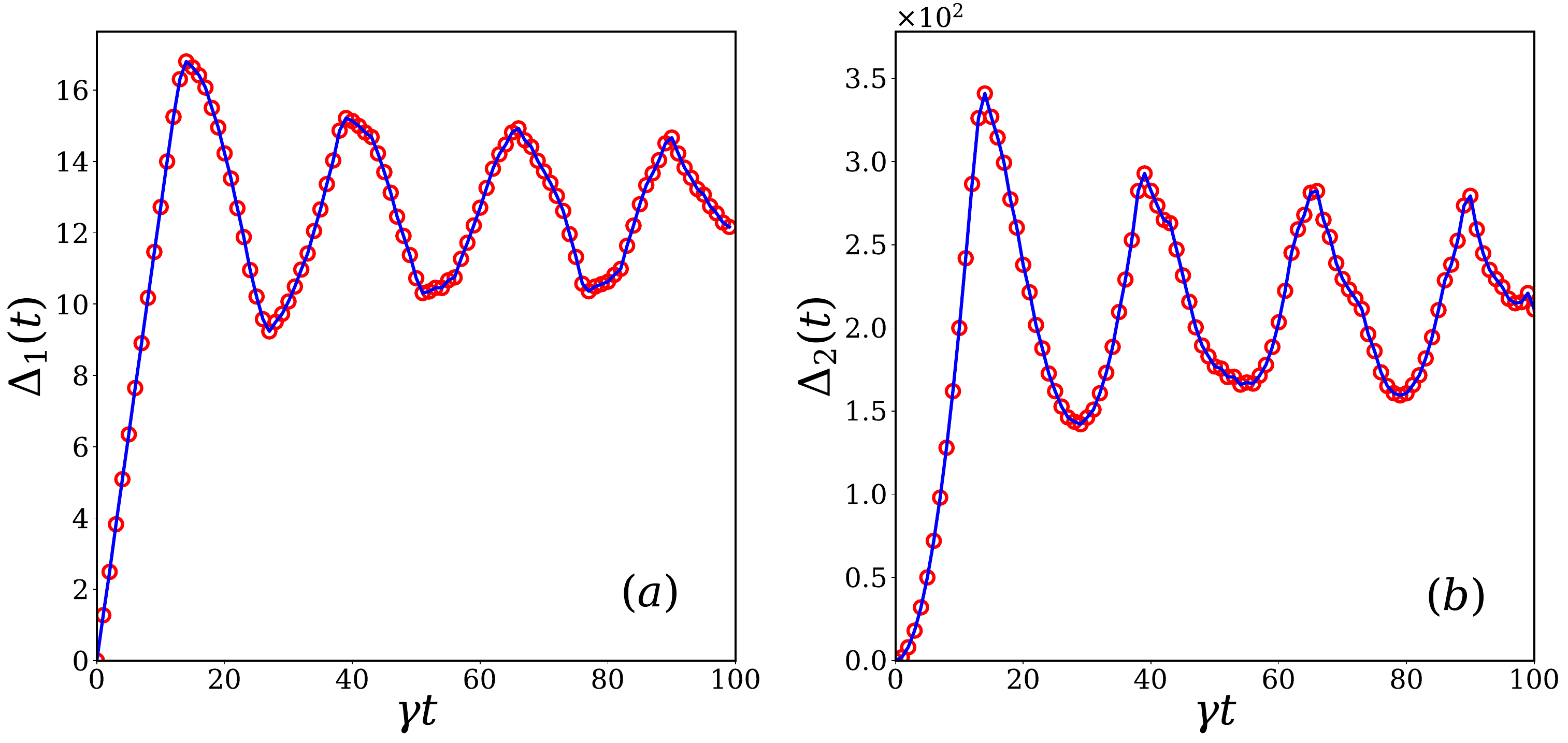}
        \caption{Defect-free mean displacement and MSD, showing agreement between theory (\ref{sec:AppA}) and numerical results from unitary evolution of the dynamics ($\gamma=1, N=50$).} \label{fig:Sup1}
\end{figure}

For $p=1$, using
\begin{align}
    \sum_{n=1}^{N-1}[n]\cos\left(\frac{2\pi y n}{N}\right)=\left\{\begin{array}{cc} 0; & \!\!y\,\,\mbox{even},N\,\,\mbox{even}, \\\!\!\!\!-\frac{1}{ \sin^2\left(\frac{\pi y}{N}\right)}; & \!\!y\,\,\mbox{odd},N\,\,\mbox{even},
    \\-\frac{1}{ 4\cos^2\left(\frac{\pi y}{2N}\right)}; &\!\!y\,\,\mbox{even},N\,\,\mbox{odd},
    \\ -\frac{1}{ 4\sin^2\left(\frac{\pi y}{2N}\right)}; &\!\!y\,\,\mbox{odd},N\,\,\mbox{odd},\end{array}\right.\label{eq:sum_n cos(2piyn/N)_formula_N}
\end{align}
and Eq.~\eqref{eq:defectfree_Delta_p_t_explicit_exp1} yields $\Delta_1(t)$ for all $N$. For $p=2$, using
\begin{align}
    \sum_{n=1}^{N-1}[n]^2\cos\left(\frac{2\pi y n}{N}\right)=\left\{\begin{array}{cc}\!\!\!\!\!\frac{N (-1)^y}{2 \sin^2\left(\frac{\pi y}{N}\right)}; & ~\forall~y,\,N\,\,\mbox{even}, 
    \\\frac{N (-1)^y \cos\left(\frac{\pi y}{N}\right)}{2 \sin^2\left(\frac{\pi y}{N}\right)} & ~\forall~y,\,N\,\,\mbox{odd},\end{array}\right.\label{eq:sum_n^2cos(2piyn/N)_formula}
\end{align}
together with Eq.~\eqref{eq:defectfree_Delta_p_t_explicit_exp1} yields an exact closed-form expression for $\Delta_2(t)$ for all $N$. Figure~\ref{fig:Sup1} shows agreement between theory and numerical results for both $\Delta_1(t)$ and~$\Delta_2(t)$. 
\vspace{-0.3cm}

%%%%%%%%%%%%%%%%%%%%%%%%%%%%%%%%%%%%%%%%%%%%%%%%%%%%%%%%%%

\section{$\Delta_1(t)$ corresponding to Fig.~\ref{fig:Fig1} of the main text}\label{sec:AppB}
\begin{figure}[h]
\centering
\includegraphics[scale=0.30]{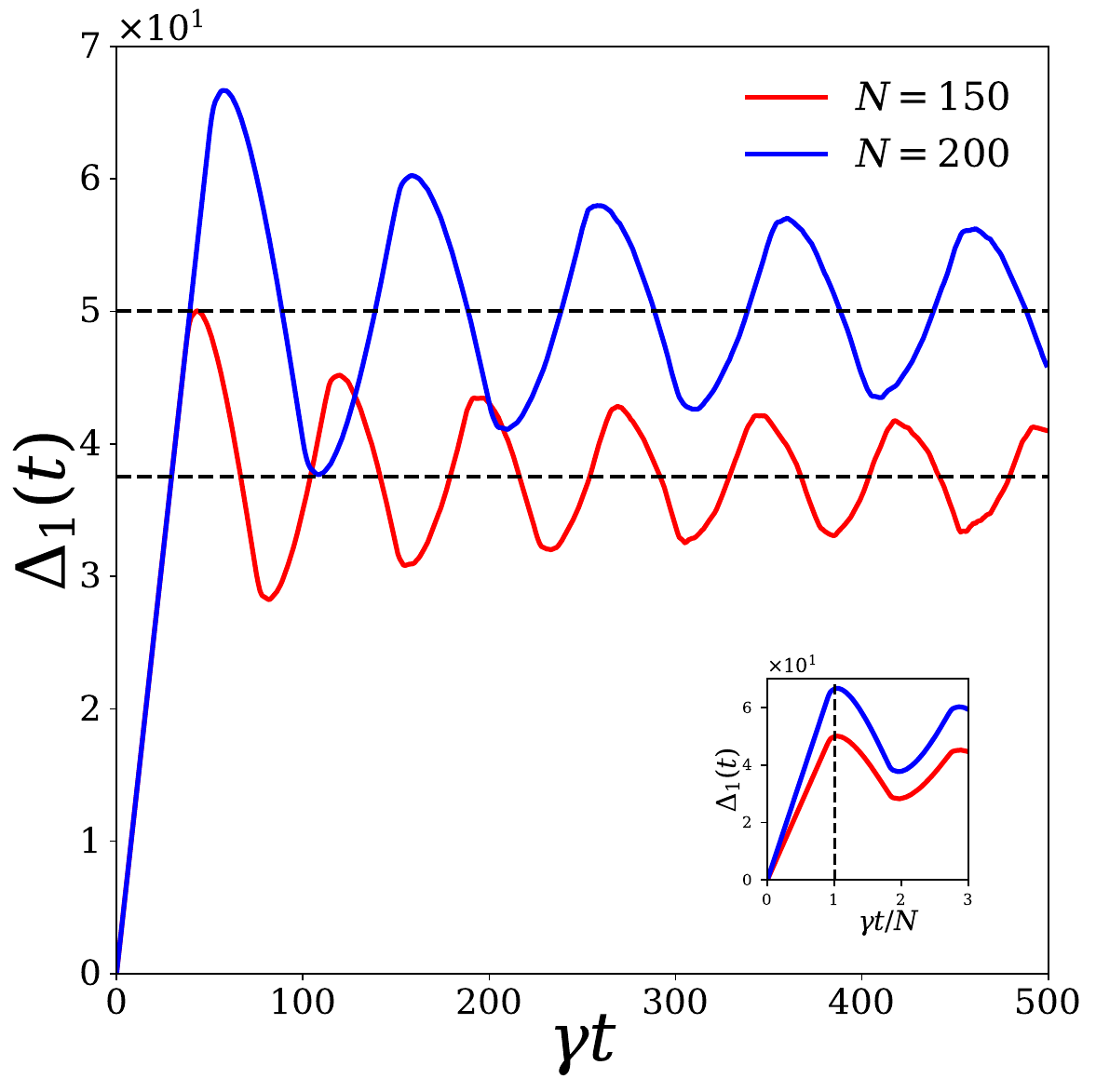}
\caption{Mean displacement $\Delta_1(t)$ versus $t$ in absence of defect, with all parameters as in  Fig.~\ref{fig:Fig1} of the main text. The dashed line denotes the steady-state value $\overline{\Delta}_1$, Eq.~\eqref{eq:mean_steady_evenN}. The timescale $t^\star \sim N/\gamma$, defined in the main text, can be seen in the inset.}
\label{fig:Fig6}
\end{figure}
%%%%%%%%%%%%%%%%%%%%%%%%%%%%%%%%%%%%%%%%%%%%%%%%%%%%%%%%%%
\vspace{-0.6 cm}

\section{Expressions of $I_n^{(d)}(t)$ and $ K_n^{(d)}(t)$}\label{sec:AppC}
Here we simplify the expression of $I_n^{(d)}(t)$ from Eq.~\eqref{eq:Probability_with_defect-1} of the main text. Straightforward  algebraic manipulation gives 
\begin{align}
    I_n^{(d)}(t)=\frac{q}{N^2}\sum_{j}\sum_{k_1,k_2}&\frac{f_j}{\mathcal{C}_{k_2}(x_j)} \Big\{\cos{\left[\beta(k_1,k_2) t\right]}\cos{\left[\mathcal{B}(n,k_1,k_2)\right]}\nonumber \\
    &+\sin\left[\beta(k_1,k_2) t\right] \sin\left[\mathcal{B}(n,k_1,k_2)\right]\nonumber\\
    &-\cos\left[2\mathcal{C}_{k_1}(x_j)t\right]\cos\left[\mathcal{B}(n,k_1,k_2)\right]
    +\sin\left[2\mathcal{C}_{k_1}(x_j)t\right]\sin\left[\mathcal{B}(n,k_1,k_2)\right]\Big\},\label{eq:Full_expression_I_nd_t}
\end{align}
with $\mathcal{B}(n,k_1,k_2) \equiv 2 \pi (k_2(n-n_d)- k_1(n-n_0))/N$, $\mathcal{C}_{k_{\alpha}}(x_j) \equiv \gamma (\cos({2 \pi k_{\alpha} /N}) -x_j )~[\alpha \in [1,2]$, with $x_j$ being the roots of $\mathcal{Q}(x)$ defined after Eq.~\eqref{eq:phi_t_q_finite}, and with $\beta(k_1,k_2)$ defined before Eq.~\eqref{Eq:msd1_time_defectfree_PBC}. Similarly, we may obtain the time-dependent expression of $K_n^{(d)}$ as
\begin{align}
   K_n^{(d)}(t)&= \frac{q^2}{4 N^2}\sum_{j,r,k_1,k_2}\frac{f_j f_r}{\mathcal{C}_{k_1}(x_j)\mathcal{C}_{k_2}(x_r)}\cos [\chi(n,n_d)]\nonumber \\
   &\times\Big[\cos({W t})+\cos({\beta t})-\cos({2 \mathcal{C}_{k_2}(x_j) t})-\cos({2 \mathcal{C}_{k_1}(x_r) t})\Big],\label{eq:Full_expression_K_nd_t}
\end{align}
with $W \equiv 2 \gamma(x_j-x_r)$.

\section{Steady state values of $I_n^{(d)}(t)$ and $K_n^{(d)}(t)$}\label{sec:AppD}
We now derive the time-average expressions of $I_n^{(d)}(t)$ starting from Eq.~\eqref{eq:Full_expression_I_nd_t}. Noting that $1/T\int_0^T dt \cos{\Theta t}=(\sin \Theta T)/\Theta T$ is non-zero in the limit $T\to \infty$ \textit{iff} $\Theta=0$, we can obtain the steady state for $I_n^{(d)}(t)$ from Eq.~\eqref{eq:Full_expression_I_nd_t} when $\beta(k_1,k_2)=0$, implying either $k_1 = k_2$ or $k_1 + k_2 = N$ with $k_1 \neq k_2$. Note that there is no contribution from $\mathcal{C}_{k_{\alpha}}(x_j)$, since $\cos (2 \pi k_\alpha/ N)- x_j \neq 0$, with $x_j$ being the roots of $\mathcal{Q}(x)$. This yields the time-average value of $I_n^{(d)}(t)$, denoted by $\overline{I}_n^{(d)}$, as
\begin{align}
    \overline{I}_n^{(d)} &= \frac{q}{N^2} \sum_{j,k_1,k_2}\frac{f_j}{\mathcal{C}_{k_2}(x_j)}\cos\left[\mathcal{B}(n,k_1,k_2)\right]\biggr|_{k_1=k_2, k_1+k_2=N}\nonumber\\
    &= \frac{q}{N^2} \left\{\sum_{j,k_1,k_2}\frac{f_j}{\mathcal{C}_{k_2}(x_j)}\cos\left[\mathcal{B}(n,k_1,k_2)\right]\biggr|_{k_1=k_2}+\sum_{j,k_1,k_2}\frac{f_j}{\mathcal{C}_{k_2}(x_j)}\cos\left[\mathcal{B}(n,k_1,k_2)\right]\biggr|_{k_1+k_2=N} \right\}\label{eq:I_d_1},
\end{align}
where in the second line, we have separated the terms corresponding to $k_1 = k_2$ and those with $k_1 + k_2 = N$ (with $k_1 \neq k_2$).  This leads to simplifying $\mathcal{B}(n,k_1,k_2)$ as
\begin{align}
\mathcal{B}=\begin{cases}
        \frac{2 \pi}{N}k(n_0-n_d)~~\forall~k_1=k_2=k, \\~
       \frac{2 \pi k}{N}(2 n-n_0-n_d)~~~\forall~~ k_2=N-k_1,
         \end{cases} 
\end{align}
which may be substituted in Eq.~\eqref{eq:I_d_1} to yield
\begin{align}
\overline{I}_n^{(d)} &= \frac{q}{N^2} \sum_{j\in \mathrm{poles}} f_j \Bigg\{\sum_{k = 0}^{N - 1} \frac{1}{\mathcal{C}_{k}(x_j)} \cos\left(\frac{2\pi k}{N}(n_0 - n_d)\right) \nonumber \\
&~~~~+  \sum_{k = 1}^{N - 1}\frac{1}{\mathcal{C}_{k}(x_j)} \cos\left(\frac{2\pi k}{N}(2n-n_0 - n_d)\right) \Bigg\}\nonumber\\
&=\frac{q}{N^2}
\sum_{j\in\mathrm{poles}} f_j \Bigg\{
\frac{1}{\mathcal{C}_0(x_j)}+2\!\sum_{k=1}^{N-1}\! 
\frac{1}{\mathcal{C}_k(x_j)} 
\cos\!\left(\frac{2\pi k}{N}(n-n_d)\right)
\!\cos\!\left(\frac{2\pi k}{N}(n-n_0)\right)
\Bigg\}.\label{eq:overline_In_d1}
\end{align}
We now turn to the steady state expression of $K_n^{(d)}(t)$, starting from Eq.~\eqref{eq:Full_expression_K_nd_t}. By the same reasoning as above, since $\mathcal{C}_{k_1}(x_r)\neq 0$ and $\mathcal{C}_{k_2}(x_j)\neq 0$, only the first two terms of Eq.~\eqref{eq:Full_expression_K_nd_t} contribute in the steady state, yielding the time-average value of $K_n^{(d)}(t)$, denoted by $\overline{K}_n^{(d)}$, as 
\begin{align}
\overline{K}_n^{(d)}\!\! &=\!\frac{q^2}{4 N^2} \!\!\Bigg\{\sum_{j\in\mathrm{poles}}\sum_{k_1,k_2=0}^{N-1} f_j^2 \frac{\cos{(\chi(n,n_d))}}{\mathcal{C}_{k_1}(x_j) \mathcal{C}_{k_2}(x_j)}\!+\!\!\! \sum_{j,r\in \mathrm{poles}} \!\Big[\sum_{k=0}^{N-1} \frac{f_j f_r}{\mathcal{C}_{k}(x_j) \mathcal{C}_{k}(x_r)}\!\nonumber \\
&+\! \sum_{k=1}^{N-1} \frac{f_j f_r}{\mathcal{C}_{k}(x_j) \mathcal{C}_{k}(x_r)}\cos{\left(\!\frac{4 \pi k(n-n_d)}{N}\!\right)}\Big]\Bigg\},\label{eq:overline_Kn_d1}
\end{align}
with $\chi(n,n_d)\equiv 2\pi (k_1-k_2) (n-n_d)/N$. Equations \eqref{eq:overline_In_d1} and \eqref{eq:overline_Kn_d1} comprise Eqs. \eqref{eq:I_n_d_qfinite} and \eqref{eq:K_n_d_qfinite} of the main text. 

\section{$\overline{I}_n^{(d)}$ and $\overline{K}_n^{(d)}$ in the limit $q\to\infty$}\label{sec:AppE}
As explained after Eq.~\eqref{eq:Phi_t_qinf_maintext1} of the main text, it has the same form as that of Eq.~\eqref{eq:phi_t_q_finite} of the main text, which was used to derive $\overline{I}_n^{(d)}$ and $\overline{K}_n^{(d)}$ in the previous section for finite $q$. Therefore, by analogy with Eq.~\eqref{eq:overline_In_d1}, we may obtain the quantity $\overline{\mathbb{I}}_n^{(d)}$, i.e.
the $q\to \infty$ counterpart of $\overline{I}_n^{(d)}$, for even $N$ as
\begin{equation}
\begin{split}
    \overline{\mathbb{I}}_n^{(d)} = -\frac{\tilde{\gamma}}{N^2} \sum_{l=1}^{N/2} F_l \Biggl[ 
    &\sum_{k=0}^{N-1} \frac{1}{\mathbb{C}_{k}(\mathrm{x}_l)} \cos \left( \frac{2\pi k}{N}(n_0 - n_d) \right) \\
    + &\sum_{k=1}^{N-1} \frac{1}{\mathbb{C}_{k}(\mathrm{x}_l)} \cos \left( \frac{2\pi k}{N}(2n - n_0 - n_d) \right) \Biggr],
\end{split}\label{eq:Ind_qinf_exp_1st}
\end{equation}
where we have $\tilde{\gamma}=4 \gamma/N$, $\mathbb{C}_{k}(\mathrm{x}_l) \equiv \gamma (\cos({2 \pi k_{\alpha} /N}) -\mathrm{x}_l )$, $\mathrm{x}_l=\cos((2l-1)\pi/N)$ with $l\in \{1,2,...,N/2\}$ for even $N$, together with $F_l\equiv\sin(|n_d-n_0| (2l-1)\pi/N)\sin( (2l-1)\pi/N)$. On simplification, we get
\begin{align}
    \overline{\mathbb{I}}_n^{(d)} &=-\frac{4}{N^3} \sum_{l=1}^{N/2} F_l \left[\frac{1}{1-\cos{\left(\frac{(2l-1)\pi}{N}\right)}}\right. \nonumber \\
    &~~~~+  \left.\sum_{k = 1}^{N - 1}\frac{\cos\left((n_0-n_d)\frac{(2l-1)\pi}{N}\right)+\cos\left((2n-n_0-n_d)\frac{(2l-1)\pi}{N}\right)}{\cos{\left(\frac{2 k \pi}{N}\right)}-\cos{\left(\frac{(2l-1)\pi}{N}\right)}} \right],\label{eq:Ind_qinf_exp2}
\end{align}
where in the last line, we used $\tilde{\gamma}=4 \gamma/N$. Now we will use the following identity from Ref.~\cite{PhysRevX.10.021045}, that
\begin{align}
    \sum_{k=1}^{N-1}\frac{\cos(\frac{2 \pi m k}{N})}{\sigma-\cos(\frac{2 \pi k}{N})}=\frac{1}{1-\sigma}+N \frac{T_{N-|m|}(\sigma)+ T_{|m|}(\sigma)}{(\sigma^2-1)U_{N-1}(\sigma)}.
\end{align}
We may define $S_m \equiv  \sum_{k = 1}^{N - 1}\cos\left( m \phi_k\right)/(\sigma- \cos(\phi_k))$, so that we get
\begin{align}
     S_m = \frac{1}{1-\sigma}+ N \frac{A_m(\sigma)}{B(\sigma)}.\label{eq:Sm_beforelimit}
\end{align}
We first analyze the behaviour of $A_m(\sigma)/B(\sigma)$ at $\sigma=\cos{\theta_l}$. Note that $A_n(x_l)= T_{N-|m|}((N-|m|)\theta_l)+T_{|m|}(|m|\theta_l)=-1+1=0$ and $U_{N-1}(\mathrm{x}_l)= \sin(N\theta_l)/\sin(\theta_l)=0$. Therefore, $A_m(\sigma)/B(\sigma)$ is of the form $0/0$, i.e.,  indeterminate, which has to be handled carefully after taking the limit $\sigma\to x_l$. We get
\begin{align}
    \lim_{\sigma\to \cos{\theta_l}}S_m = \frac{1}{1-\cos{\theta_l}}+N \frac{A'_m(\sigma)}{B'(\sigma)}\bigg|_{\sigma=\cos{\theta_l}},
\end{align}
with the primes denoting first-order derivatives with respect to $\sigma$. Assuming $\sigma= \cos(\theta)$, for any function $f(\sigma)=\tilde{f}(\theta)$, we have using chain rule of differentiation that $d f /d\theta|_{\sigma=\cos{\theta}} = - (1/\sin{\theta}) d \tilde{f}/d \theta$, which may be used to obtain
\begin{align}
    \frac{A'_m(\sigma)}{B'(\sigma)}
    &= 
    \frac{
      \left.\frac{d}{d\theta}\!\left[\cos\big((N-|m|)\theta\big)+\cos(|m|\theta)\right]\right|_{\theta=\theta_l}}{\left.\frac{d}{d\theta}\left[-\sin^2\theta\, U_{N-1}(\cos\theta)\right]\right|_{\theta=\theta_l}
    }=-\frac{\sin\!\big(|m|\theta_l\big)}{\sin\theta_l}.
\end{align}
Thus, we obtain from Eq.~\eqref{eq:Sm_beforelimit} that
\begin{align}
    \lim_{\sigma\to \cos{\theta_l}}S_m = \frac{1}{1-\cos{\theta_l}}-N \frac{\sin{|m|\theta_l}}{\sin{\theta_l}}.\label{eq:Sm_postlimit}
\end{align}
Now, using Eq.~\eqref{eq:Sm_beforelimit} in Eq.~\eqref{eq:Ind_qinf_exp2}, we get (using $F_l=\sin{|n_0-n_d| \theta_l} \sin{\theta_l}$)
\begin{align}
    \overline{\mathbb{I}}_n^{(d)} 
    &=\frac{4}{N^3} \sum_{l=1}^{N/2}  \Big( \frac{\sin{\left(|n_0-n_d| \theta_l\right)}\sin{\theta_l}}{1-\cos{\theta_l}}-  N  \sin^2{(|n_0-n_d|\theta_l)} \nonumber \\
    &~~~~- N \sin{(|n_0-n_d|\theta_l)}\sin{|2 n -n_0-n_d|\theta_l} \Big).\label{eq:Ind_inf_exp3}
\end{align}

Now we will evaluate the three sums in Eq.~\eqref{eq:Ind_inf_exp3} explicitly. We proceed to evaluate the first term in Eq.~\eqref{eq:Ind_inf_exp3} as follows: 
\begin{align}
\sum_{l=1}^{N/2}\frac{\sin\left(|n_d-n_0|\theta_l\right)\sin{\theta_l}}{1-\cos{\theta_l}}=\frac{1}{2}\left\{\sum_{l=1}^{N/2}\frac{\cos\left(\big(|n_d-n_0|-1\big)\theta_l\right)}{1-\cos{\theta_l}}-\sum_{l=1}^{N/2}\frac{\cos\left(\big(|n_d-n_0|+1\big)\theta_l\right)}{1-\cos{\theta_l}}\right\}.\label{eq:Ind_inf_1stterm1}
 \end{align}
We can show for any integer $m$ (with $\theta_l = (2l-1)\pi/N$) that $\sum_{l=1}^{N/2}(\cos(m\theta_l))/(1-\cos(\theta_l))=N/2\left(N/2-|m|\right)$,
which, after using in Eq.~\eqref{eq:Ind_inf_1stterm1}, we get
\begin{align}
&\sum_{k=1}^{N/2}\frac{\sin\left[|n_d-n_0|\frac{(2k-1)\pi}{N}\right]\sin\left[\frac{(2k-1)\pi}{N}\right]}{1-\cos\left[\frac{(2k-1)\pi}{N}\right]}\nonumber \\
&=\frac{1}{2}\left[\frac{N}{2}\left(\frac{N}{2}+1-|n_d-n_0|\right)-\frac{N}{2}\left(\frac{N}{2}-1-|n_d-n_0|\right)\right]=\frac{N}{2},
\end{align}
so that
\begin{align}
\sum_{l=1}^{N/2}\frac{\sin\left(|n_d-n_0|\theta_l\right)\sin{\theta_l}}{1-\cos{\theta_l}}=\frac{N}{2}(1-\delta_{n_d,n_0}).
\end{align}
For the second term in Eq.~\eqref{eq:Ind_inf_exp3}, we may obtain
\begin{align}
    \sum_{l=1}^{N/2}\sin^2\left(|n_d-n_0|\theta_l\right)=\frac{N}{4}(1-\delta_{n_d,n_0}),
\end{align}
and that for the third term in Eq.~\eqref{eq:Ind_inf_exp3}, we have
\begin{align}
&\sum_{l=1}^{N/2}\sin\left(|n_d-n_0|\theta_l\right)\sin\left(|2n-n_d-n_0|\theta_l\right)\nonumber \\
&=\frac{N}{4}(1-\delta_{n_0,n_d}\delta_{2n,n_0+n_d})\left(\delta_{|2n-n_d-n_0|,|n_d-n_0|}-\delta_{n,n_0}\delta_{n_0,n_d}\right).
\end{align}
Using the expressions of the all three terms in Eq.~\eqref{eq:Ind_inf_exp3}, we obtain
\begin{align} \overline{\mathbb{I}}_n^{(d)}&=\frac{4}{N^3}\biggl(\frac{N}{2}(1-\delta_{n_d,n_0})- \frac{N^2}{4}(1-\delta_{n_d,n_0})\nonumber \\
&~~~~-\frac{N^2}{4}(1-\delta_{n_0,n_d}\delta_{2n,n_0+n_d})\left(\delta_{|2n-n_d-n_0|,|n_d-n_0|}-\delta_{n,n_0}\delta_{n_0,n_d}\right) \biggr)\nonumber\\
    &=-\frac{N-2}{N^2}(1-D)-\frac{1}{N} (U-V D-D E U+ V E D^2),
\end{align} 
where we have defined the Kronecker delta expressions $D\equiv\delta_{n_d,n_0}$, $E\equiv\delta_{2n,n_0+n_d}$, $U\equiv\delta_{|2n-n_d-n_0|,|n_d-n_0|}$ and $V\equiv\delta_{n,n_0}$. Now since we are interested in $n_d\neq n_0$, we have $D=0$, which yields for even $N$
\begin{align}
\overline{\mathbb{I}}_n^{(d)}&=-\frac{N-2}{N^2}-\frac{1}{N} \delta_{|2n-n_d-n_0|,|n_d-n_0|}\nonumber\\
&=-\frac{N-2}{N^2}-\frac{1}{N} \left(\delta_{n,n_0}+\delta_{n,n_d}+\delta_{n,n_0+\frac{N}{2}}+\delta_{n,n_d+\frac{N}{2}}\right),\label{eq:Ind_qinf_exp4}
\end{align} 
which is the first part of Eq.~\eqref{eq:I_n_d_qinfinite_even_odd_N} of the main text. Using Eq.~\eqref{eq:Ind_qinf_exp4}, one obtains for $n=n_0, n_d, n_0+N/2, n_d+ N/2$ that $\overline{\mathbb{I}}_{n_0}^{(d)}=\overline{\mathbb{I}}_{n_d}^{(d)}=\overline{\mathbb{I}}_{n_0+\frac{N}{2}}^{(d)}=\overline{\mathbb{I}}_{n_d+\frac{N}{2}}^{(d)}$. Now for odd $N$, we start with Eq.~\eqref{eq:Ind_inf_exp3}, with $l\in \{1,2,...,(N-1)/2\}$ for odd $N$ and get
\begin{align}
    \overline{\mathbb{I}}_n^{(d)} &=\frac{4}{N^3} \sum_{l=1}^{(N-1)/2}  \Big( \frac{\sin{\left(|n_0-n_d| \theta_l\right)}\sin{\theta_l}}{1-\cos{\theta_l}}-  N  \sin^2{(|n_0-n_d|\theta_l)} \nonumber \\
    &~~~~- N \sin{(|n_0-n_d|\theta_l)}\sin{|2 n -n_0-n_d|\theta_l} \Big).\label{eq:Ind_inf_exp5}
\end{align}
Here also, one obtains
\begin{align}
\sum_{l=1}^{(N-1)/2}\frac{\sin\left(|n_d-n_0|\theta_l\right)\sin{\theta_l}}{1-\cos{\theta_l}}=\frac{N}{2}(1-\delta_{n_d,n_0}), 
\end{align}
\begin{align}
\sum_{k=1}^{(N-1)/2}\sin^2\left[|n_d-n_0|\frac{(2k-1)\pi}{N}\right]=\frac{N}{4}(1-\delta_{n_d,n_0}), 
\end{align}
\begin{align}
    &\sum_{l=1}^{(N-1)/2}\sin\left(|n_d-n_0|\theta_l\right)\sin\left(|2n-n_d-n_0|\theta_l\right)\nonumber \\
    &=\frac{N}{4}(1-\delta_{n_0,n_d}\delta_{2n,n_0+n_d})\left(\delta_{|2n-n_d-n_0|,|n_d-n_0|}-\delta_{n,n_0}\delta_{n_0,n_d}\right),
\end{align}
which finally yields for odd $N$
\begin{align}
\overline{\mathbb{I}}_n^{(d)}&=-\frac{N-2}{N^2}-\frac{1}{N}\left( \delta_{n,n_0}+\delta_{n,n_d}\right).\label{eq:Ind_qinf_exp4_oddN}
\end{align}
Using Eq.~\eqref{eq:Ind_qinf_exp4_oddN}, one obtains for $n=n_0, n_d$ that $\overline{\mathbb{I}}_{n_0}^{(d)}=\overline{\mathbb{I}}_{n_d}^{(d)}$ for odd $N$. Equation \eqref{eq:Ind_qinf_exp4_oddN} matches with the second part of Eq.~\eqref{eq:I_n_d_qinfinite_even_odd_N} of the main text.

Now we proceed with the quantity $\overline{\mathbb{K}}_n^{(d)}$, i.e. the $q\to \infty$ counterpart of $\overline{K}_n^{(d)}$ and get
\begin{align}
&\overline{\mathbb{K}}_n^{(d)} =\frac{4 {\gamma}^2}{N^4} \Biggl[\sum_{l=1}^{\lfloor N/2 \rfloor}\sum_{k_1,k_2=0}^{N-1} \frac{F_l^2}{\mathbb{C}_{k_1}(\mathrm{x}_l) \mathbb{C}_{k_2}(\mathrm{x}_l)}  \cos(\chi(n,n_d))\nonumber\\&+\sum_{l,m=1}^{\lfloor N/2 \rfloor}\left( \sum_{k=0}^{N-1}  \frac{F_l F_m}{\mathbb{C}_{k}(\mathrm{x}_l) \mathbb{C}_{k}(\mathrm{x}_m)} +\sum_{k=1}^{N-1} \frac{F_l F_m}{\mathbb{C}_{k}(\mathrm{x}_l) \mathbb{C}_{k}(\mathrm{x}_m)}  \cos{(4 \pi k (n-n_d)/N)} \right)\Biggr],
\end{align}
with $\lfloor N/2 \rfloor = N/2$ for even $N$ and $\lfloor N/2 \rfloor = (N-1)/2$ for odd $N$. We start with 
\begin{align}
    &\sum_{k_1,k_2=0}^{N-1} \frac{F_l^2}{\mathbb{C}_{k_1}(\mathrm{x}_l) \mathbb{C}_{k_2}(\mathrm{x}_l)}  \cos(2\pi (k_1-k_2) (n-n_d)/N)\nonumber \\
    &=F_l^2~\mathbf{Re}\left[\sum_{k_1,k_2=0}^{N-1} \frac{1}{\mathbb{C}_{k_1}(\mathrm{x}_l)\mathbb{C}_{k_2}(\mathrm{x}_l)}  e^{2i\pi (k_1-k_2) (n-n_d)/N}\right]\nonumber\\
    &=F_l^2~\left|\sum_{k=0}^{N-1} \frac{e^{2i\pi k(n-n_d)/N}}{\mathbb{C}_{k}(\mathrm{x}_l)}  \right|^2,
\end{align}
where $\mathbf{Re}$ stands for ``real part of." Defining $R_k\equiv \sum_{m=1}^{\lfloor N/2 \rfloor} \frac{F_m}{\mathbb{C}_{k}(\mathrm{x}_m)}$, we may write
\begin{align}
    \sum_{l,m=1}^{\lfloor N/2 \rfloor} \sum_{k=0}^{N-1}  \frac{F_l F_m}{\mathbb{C}_{k}(\mathrm{x}_l) \mathbb{C}_{k}(\mathrm{x}_m)}&=\sum_{k=0}^{N-1} R_k ^2,
\end{align}
and 
\begin{align}
\sum_{l,m=1}^{\lfloor N/2 \rfloor} \sum_{k=0}^{N-1}  \frac{F_l F_m}{\mathbb{C}_{k}(\mathrm{x}_l) \mathbb{C}_{k}(\mathrm{x}_m)} \cos{(4 \pi k (n-n_d)/N)}&=\sum_{k=0}^{N-1} R_k ^2 \cos{(4 \pi k (n-n_d)/N)},
\end{align}
to finally yield
\begin{align}
&\overline{\mathbb{K}}_n^{(d)} =\frac{4 {\gamma}^2}{N^4} \Biggl[\sum_{l=1}^{\lfloor N/2 \rfloor}F_l^2~\left|\sum_{k=0}^{N-1} \frac{e^{2i\pi k(n-n_d)/N}}{\mathbb{C}_{k}(\mathrm{x}_l)}  \right|^2+\sum_{k=0}^{N-1} R_{k}^2+\sum_{k=1}^{N-1} R_{k}^2 \cos{(4 \pi k (n-n_d)/N)}\Biggr],\label{eq:Knd_qinf_exp1}
\end{align}
which is Eq.~\eqref{eq:K_n_d_qinfinite_even_odd_N} of the main text. Using Eq.~\eqref{eq:Knd_qinf_exp1}, one gets for $n=n_0, 2n_d-n_0$ that $\overline{\mathbb{K}}_{n_0}^{(d)}=\overline{\mathbb{K}}_{2n_d-n_0}^{(d)}$,
and also $\overline{\mathbb{K}}_{n}^{(d)}$ is maximum when $n=n_d+N/2$.

\section{Reflecting wall at the defect site $n_d$}\label{sec:reflecting}
This section aims to consider whether the effects in the limit $q\to \infty$ reported in Sec.~\ref{subsec:q_infinity} may be reproduced by placing at the defect site a perfectly reflecting wall. In this case, one has to consider the system with open boundaries, with the Hamiltonian given by
\begin{align}
H=-\sum_{n=0}^{N-3} \gamma\left( |n +1 \rangle  \langle n | + |n \rangle  \langle n+1 | \right).
\label{eq:hamiltonian-1}
\end{align}
In the framework of the Hamiltonian~\eqref{eq:hamiltonian}, the above one is obtained by placing the reflecting wall at the site $n=N-1$, so that there is no hopping between sites $N-2$ and $N-1$ and between sites $0$ to $N-1$. Corresponding to the Hamiltonian~\eqref{eq:hamiltonian-1}, the eigenvalues and eigenvectors are respectively given by $E_\alpha=-2\gamma \cos(\alpha \pi/N)$ and $|\phi_\alpha\rangle=\sqrt{2/N}\sum_{n=0}^{N-2}\sin(\pi \alpha(n+1)/N)|n\rangle$, with $\alpha=1,2,\ldots,N-1$~\cite{Dhar_2015}. Using Eq.~\eqref{eq:psi-01}, we have
\begin{align}
    G(n,n_0,t)&=\sum_\alpha e^{-i E_\alpha t} \phi_\alpha (n)~[\phi_\alpha(n_0)]^\star\nonumber \\
    &=\frac{2}{N}
\sum_{\alpha=1}^{N-1}
e^{2 i \gamma t \cos k_\alpha}
\sin\!\big(k_\alpha (n+1)\big)
\sin\!\big(k_\alpha (n_0 + 1)\big),
    \label{eq:psi-02}
\end{align}
yielding
%\label{eq:P_n_bar_reflecting_wall}
\begin{align}
P_n(t)
&=
\frac{4}{N^2}
\sum_{\alpha=1}^{N-1} \sum_{\alpha'=1}^{N-1}
e^{2 i \gamma t \left(\cos k_\alpha - \cos k_{\alpha'}\right)}
\sin\!\big(k_\alpha (n+1)\big)
\sin\!\big(k_\alpha (n_0 + 1)\big)\nonumber \\
&\times\sin\!\big(k_{\alpha'} (n+1)\big)
\sin\!\big(k_{\alpha'} (n_0 + 1)\big),
\end{align}
with $k_\alpha \equiv \alpha\pi/N$. We have $\sum_{n=0}^{N-2}P_n(t)=1$, where we have used the relations $\sum_{m=1}^{N-1}\sin(k_\alpha m)\sin (k_{\alpha'} m)=N/2\,\delta_{\alpha\alpha'}$ and $\sum_{\alpha=1}^{N-1}\sin^2(k_\alpha(n_0+1))=N/2$.
In the steady state, one has
\begin{align}
\overline{P_n}
&=\lim_{T \to \infty} \frac{1}{T} \int_0^T dt~P_n(t)\nonumber\\
&=\frac{4}{N^2}
\sum_{\alpha=1}^{N-1}
\sin^2\!\big(k_\alpha (n+1)\big)\,
\sin^2\!\big(k_\alpha(n_0+1)\big)\nonumber\\
&=\frac{1}{N^2}
\sum_{\alpha=1}^{N-1}
\left(1-\cos\!\big(2 k_\alpha (n+1)\big)\right)\left(1-\cos\!\big(2 k_\alpha (n_0+1)\big)\right)\nonumber\\
&=\frac{1}{N^2}
\sum_{\alpha=1}^{N-1}
\biggl(1-\cos\!\big(\frac{2 \pi \alpha}{N} (n+1)\big)-\cos\!\big(\frac{2 \pi \alpha}{N} (n_0+1)\big)+\frac{1}{2}\cos\!\big(\frac{2 \pi \alpha}{N} (n-n_0)\big)\nonumber\\
&~~+\frac{1}{2}\cos\!\big(\frac{2 \pi \alpha}{N} (n+n_0+2)\big)\biggr).\label{eq:P_n_bar_reflecting2}
\end{align}

Now, we will use the relation $\sum_{\alpha=1}^{N-1} \cos\left(2\pi \alpha \mathrm{X}/N\right) = N (\sum_{p=-\infty}^{\infty} \delta_{\mathrm{X}, p N}) - 1$.  Given that the site indices $n,n_0 \in [0, N-2]$, we have $n+1$ and $n_0+1$ that do not ever satisfy the relation $X = pN$. However, $n-n_0$ and $n+n_0+2$ satisfy the aforementioned relation for $p=0$ and $p=1$, respectively. Equation~\eqref{eq:P_n_bar_reflecting2} simplifies into
\begin{align}
    \overline{P}_n
&=\frac{1}{N}+\frac{1}{2N} \left( \delta_{n,n_0}+\delta_{n,N-n_0-2;~\mathrm{mod}~N}\right).
\end{align}
Given the fact that $\overline{P}_n=0$ for $n=N-1$, by construction, we may rewrite the last equation as
\begin{align}
\overline{P}_n&=\frac{1}{N}+\frac{1}{2N} \left( \delta_{n,n_0}+\delta_{n,N-n_0-2;~\mathrm{mod}~N}\right)-\frac{1}{N}\delta_{n,N-1}.\label{eq:P_n_steady_reflecting_defect_N-1}
\end{align}
On the other hand, Eq.~\eqref{eq:Pnd-qinf} from Sec.~\ref{subsec:q_infinity} may be simplified as
\begin{align}
\overline{P}_n&=\frac{1}{N}+\frac{1}{2N} \left( \delta_{n,n_0}+\delta_{n,2 n_d-n_0}\right)-\frac{1}{N}\delta_{n,n_d},
\end{align}
which, after putting $n_d=N-1$ and using $\delta_{n,2 N-2-n_0}=\delta_{n, N-n_0-2;~\mathrm{mod}~N}$ for a periodic chain, matches exactly with Eq.~\eqref{eq:P_n_steady_reflecting_defect_N-1}. Thus, considering the model~\eqref{eq:hamiltonian} for finite $q$ and taking the limit $q\to \infty$ yield identical results with respect to those obtained by placing an infinite wall at the defect site and considering instead the model~\eqref{eq:hamiltonian-1}. 

\section{The case of $2$ defects}\label{sec:AppF} 
Here, we sketch the derivation of the wave function in the presence of two onsite defects. As detailed after Eq.~\eqref{eq:wave_function_laplace_Ndefects} of the main text, the wave-function in the Laplace domain can be written as a matricial equation $\mathbf{M\Psi}=\mathbf{G}$, where 
\begin{align}
     \mathbf{M}=\begin{pmatrix}
  1-i q_1 \widetilde{G}(n_{d_1},n_{d_1},\epsilon)& -i q_2 \widetilde{G}(n_{d_1},n_{d_2},\epsilon)\\
    -i q_1 \widetilde{G}(n_{d_2},n_{d_1},\epsilon) &1-i q_2 \widetilde{G}(n_{d_2},n_{d_2},\epsilon)
\end{pmatrix}.
\end{align}
Now the inverse of $\mathbf{M}$ is
\begin{align}
     \mathbf{M}^{-1}=\frac{1}{\mathrm{det}{\mathbf{M}}}\begin{pmatrix}
 1-i q_2 \widetilde{G}(n_{d_2},n_{d_2},\epsilon)& i q_2 \widetilde{G}(n_{d_1},n_{d_2},\epsilon)\\
    i q_1 \widetilde{G}(n_{d_2},n_{d_1},\epsilon) &1-i q_1 \widetilde{G}(n_{d_1},n_{d_1},\epsilon)
\end{pmatrix},
\end{align}
where the determinant may be simplified using Eq.~\eqref{eq:homogeneous_G_laplace} of the main text (defining $x\equiv \epsilon/ (2i \gamma)$), yielding
\begin{align}\det \mathbf{M}&=1-\frac{q_1+q_2}{2\gamma}\frac{T_{N}(x)+1}{\left[(x)^2-1\right]U_{N-1}(x)}\nonumber \\
&-\frac{q_1 q_2}{4\gamma^2}\frac{\left[T_{N-|n_{d_1}-n_{d_2}|}(x)+T_{|n_{d_1}-n_{d_2}|}(x)\right]^2-\left[T_{N}(x)+1\right]^2}{\left\{\left[(x)^2-1\right] U_{N-1}(x)\right\}^2},
\end{align}
using which we may obtain  $\mathbf{\Psi}= \mathbf{M}^{-1}\mathbf{G}$, yielding
\begin{align}
    &\widetilde{\psi}(n_{d_1},n_0,\epsilon)=\left(\Big[1-iq_2\widetilde{G}(n_{d_2},n_{d_2},\epsilon)\Big]\widetilde{G}(n_{d_1},n_0,\epsilon)+iq_2\widetilde{G}(n_{d_1},n_{d_2},\epsilon)\widetilde{G}(n_{d_2},n_0,\epsilon)\right)\nonumber \\
    &~~~~~~~~~~~~~~~~~~~~\times\Big[\det \mathbf{M}\Big]^{-1},\nonumber\\ \\
    &\widetilde{\psi}(n_{d_2},n_0,\epsilon)=\left(\Big[1-iq_1\widetilde{G}(n_{d_1},n_{d_1},\epsilon)\Big]\widetilde{G}(n_{d_2},n_0,\epsilon)+iq_1\widetilde{G}(n_{d_2},n_{d_1},\epsilon)\widetilde{G}(n_{d_1},n_0,\epsilon)\right)\nonumber \\
    &~~~~~~~~~~~~~~~~~~~~\times \Big[\det \mathbf{M}\Big]^{-1}. \nonumber
\end{align}
Thus, we have the complete expression of wave function in the Laplace domain in the presence of $2$ defects, given by
\begin{align}
&    \widetilde{\psi}(n,n_0,\epsilon) = \widetilde{G}(n,n_0,\epsilon)+ i q_1 \widetilde{G}(n,n_{d_1},\epsilon)\widetilde{\psi}(n_{d_1},n_0,\epsilon) + i q_2 \widetilde{G}(n,n_{d_2},\epsilon) \widetilde{\psi}(n_{d_2},n_0,\epsilon) \nonumber \\
& = \widetilde{G}(n,n_0,\epsilon)+i\Big[q_1\widetilde{G}(n,n_{d_1},\epsilon)\widetilde{G}(n_{d_1},n_0,\epsilon)+q_2\widetilde{G}(n,n_{d_2},\epsilon)\widetilde{G}(n_{d_2},n_0,\epsilon)\Big]\Big[\det \mathbf{M}\Big]^{-1}\nonumber \\
&+q_1q_2\Big\{\Big[\widetilde{G}(n_{d_2},n_{d_2},\epsilon)\widetilde{G}(n_{d_1},n_0,\epsilon)-\widetilde{G}(n_{d_1},n_{d_2},\epsilon)\widetilde{G}(n_{d_2},n_0,\epsilon)\Big]\widetilde{G}(n,n_{d_1},\epsilon)\nonumber \\
&+\Big[\widetilde{G}(n_{d_1},n_{d_1},\epsilon)\widetilde{G}(n_{d_2},n_0,\epsilon)-\widetilde{G}(n_{d_2},n_{d_1},\epsilon)\widetilde{G}(n_{d_1},n_0,\epsilon)\Big]\widetilde{G}(n,n_{d_2},\epsilon)\Big\}\Big[\det \mathbf{M}\Big]^{-1},
\end{align}
which after some algebraic manipulation gives the total $\widetilde{\psi}(n,n_0,\epsilon)$ in the Laplace domain as 
\begin{align}
&\widetilde{\psi}(n,n_0,\epsilon) = \widetilde{G}(n,n_0,\epsilon)\nonumber \\
 &+\frac{1}{Q(x)}\left\{ i q_1 R_{n_{d_1}}(x)+iq_2 R_{n_{d_2}}(x)+q_1 q_2 [P_{n_{d_1}}(x) \widetilde{G}(n,n_{d_1},\epsilon)+P_{n_{d_2}}(x) \widetilde{G}(n,n_{d_2},\epsilon)]\right\}, \label{eq:psi_2defects_laplace_appendix}
\end{align}
which is Eq.~\eqref{eq:psi_2defects_laplace} of the main text. The following quantities have been defined
\begin{align}
Q(x)&\equiv\Big[(x^2-1)U_{N-1}(x)\Big]^2-\frac{q_1+q_2}{2\gamma}(x^2-1)U_{N-1}(x)\Big[T_N(x)+1\Big] \nonumber\\
& -\frac{q_1q_2}{4\gamma^2}\left\{\Big[T_{N-|n_{d_1}-n_{d_2}|}(x)+T_{|n_{d_1}-n_{d_2}|}(x)\Big]^2-\Big[T_{N}(x)+1\Big]^2\right\},
\end{align}
and
\begin{align}
R_{\alpha}(x)&\equiv\frac{1}{4\gamma^2}\Big[T_{N-|\alpha-n|}\left(x\right)+T_{|\alpha-n|}\left(x\right)\Big]\Big[T_{N-|\alpha-n_0|}\left(x\right)+T_{|\alpha-n_0|}\left(x\right)\Big],
\end{align} and

\begin{align}
    P_{n_{d_1}}(x)&\equiv \Big[T_{N-|n_{d_1}-n_0|}\left(x\right)+T_{|n_{d_1}-n_0|}\left(x\right)\Big]\Big[T_N(x)+1\Big] \nonumber\\
    &-\Big[T_{N-|n_{d_1}-n_{d_2}|}\left( x\right)+T_{|n_{d_1}-n_{d_2}|}\left(x\right)\Big]\Big[T_{N-|n_{d_2}-n_0|}\left(x\right)+T_{|n_{d_2}-n_0|}\left(x\right)\Big]\nonumber \\
P_{n_{d_2}}(x)&\equiv\Big[T_{N-|n_{d_2}-n_0|}\left(x\right)+T_{|n_{d_2}-n_0|}\left(x\right)\Big]\Big[T_N(x)+1\Big] \nonumber\\&-\Big[T_{N-|n_{d_1}-n_{d_2}|}\left( x\right)+T_{|n_{d_1}-n_{d_2}|}\left(x\right)\Big]\Big[T_{N-|n_{d_1}-n_0|}\left(x\right)+T_{|n_{d_1}-n_0|}\left(x\right)\Big].
\end{align}
Upon Laplace inverse transformation, Eq~\eqref{eq:psi_2defects_laplace_appendix} yields the time-dependent wave function in the presence of $2$ defects. 

\section{A classical model of a random walker on a one-dimensional lattice with a symmetric, partially permeable barrier between two sites}\label{sec:AppG} 
Here we will discuss the case of a continuous-time random walker on a one-dimensional lattice with a symmetric, partially permeable barrier between two sites~\cite{PhysRevResearch.4.L032039}. The occupation probability 
$P(n,t)$ evolve according to the master equation
\begin{align}
\frac{dP(n,t)}{dt}&=F\Big[P(n+1,t)+P(n-1,t)-2P(n,t)\Big]\nonumber \\
&-\Delta \Big[P(r+1,t)-P(r,t)\Big]\Big(\delta_{n,r}-\delta_{n,r+1}\Big),
\end{align}
where $\Delta=F-f$ encodes the reduced hopping rate with $f<F$ across the barrier between sites $r$ and $r+1$. The defect technique gives the exact Laplace solution
\begin{align}
    &\widetilde{P}_{n_0}(n,\epsilon)=\widetilde{\Psi}_{n_0}(n,\epsilon)\nonumber \\
    &-\Big[\widetilde{\Psi}_{r}(n,\epsilon)-\widetilde{\Psi}_{r+1}(n,\epsilon)\Big]\frac{\widetilde{\Psi}_{n_0}(r+1,\epsilon)-\widetilde{\Psi}_{n_0}(r,\epsilon)}{\frac{1}{\Delta}+\widetilde{\Psi}_{r+1}(r,\epsilon)+\widetilde{\Psi}_{r}(r+1,\epsilon)-\widetilde{\Psi}_{r+1}(r+1,\epsilon)-\widetilde{\Psi}_{r}(r,\epsilon)},
\end{align}
with $\widetilde{\Psi}_{n_0}(n,\epsilon)$ given by
\begin{align}
\widetilde{\Psi}_{n_0}(n,\epsilon)=\frac{1}{N\epsilon}+\frac{1}{N}\sum_{k=1}^{N-1}\frac{\cos\left[(n-n_0)\frac{2\pi k}{N}\right]}{\epsilon+2F\left[1-\cos\left(\frac{2\pi k}{N}\right)\right]}.
\end{align}
Now the MSD is given by $S(t)=\sum_{n=0}^{N-1}(n-n_0)^2 P_n (t)$. The steady-state value of 
$S(t)$ can be determined via $\widetilde{\Psi}_{n_0}(n,\epsilon)$, whose steady state value can be obtained using the final value theorem, which reads as $\lim_{t\to\infty} S (t) =\lim_{\epsilon\to0}\epsilon \widetilde{S}(\epsilon)$. Clearly, the effect of the defect strength (here, the reduced hopping strength) is suppressed upon evaluating the $\epsilon\to0$ limit and hence the steady state value of $S(t)$ is independent of $\Delta$, in contrast to the results discussed in the main text.
\end{document}